\documentclass[lettersize,journal]{IEEEtran}
\usepackage{amsmath,amsfonts}
\usepackage{algorithm}
\usepackage{array}
\usepackage[caption=false,font=footnotesize]{subfig}
\usepackage[rightcaption]{sidecap} 
\sidecaptionvpos{figure}{c}        

\usepackage{textcomp}
\usepackage{stfloats}
\usepackage{url}
\usepackage{verbatim}
\usepackage{graphicx}
\usepackage{cite}
\newcolumntype{P}[1]{>{\centering\arraybackslash}p{#1}}
\usepackage{colortbl}

\usepackage{caption}
\usepackage{subcaption} 

\usepackage{xr}
\usepackage{amsmath,amsfonts}
\usepackage{graphicx}
\usepackage{textcomp}
\usepackage{xcolor}
\usepackage{comment}
\usepackage{booktabs}
\usepackage{enumitem}
\usepackage[caption=false]{subfig}
\usepackage{url}
\usepackage[normalem]{ulem}
\usepackage{algorithm}
\usepackage{algpseudocode}
\usepackage{array}
\usepackage{newtxmath}
\usepackage{euscript} 
\usepackage[numbers]{natbib}
\usepackage{wrapfig}
\usepackage{hyperref}
\usepackage{amsmath}

\usepackage{tikz}
\usetikzlibrary{arrows.meta,positioning,shadows.blur,calc}

\begin{document}

\title{Assessing Socio-Cyber Vulnerability Using Survey and Social Media Data}

\author{Shutonu~Mitra,
        Qi~Zhang,
        Tomas~Neguyen,        
        Hossein~Salemi,
        Fengxiu~Zhang,
        Michin~Hong,
        Chang-Tien~Lu,~\IEEEmembership{Fellow,~IEEE,}
        Hemant~Purohit,
        Jin-Hee~Cho,~\IEEEmembership{Senior~Member,~IEEE,}%
\thanks{Shutonu Mitra (email: \href{mailto:mshutonu@vt.edu}{mshutonu@vt.edu}), Qi Zhang (email: \href{mailto:qiz21@vt.edu}{qiz21@vt.edu}), Tomas Neguyen (email: \href{mailto:thomasn03@vt.edu}{thomasn03@vt.edu}), Chang-Tien Lu (email: \href{mailto:clu@vt.edu}{clu@vt.edu}), and Jin-Hee Cho (email: \href{mailto:jicho@vt.edu}{jicho@vt.edu}) are with Virginia Tech, Blacksburg, VA, USA.}%
\thanks{Hossein Salemi (email: \href{mailto:hsalemi@gmu.edu}{hsalemi@gmu.edu}), Fengxiu Zhang (email: \href{mailto:fzhang22@gmu.edu}{fzhang22@gmu.edu}), and Hemant Purohit (email: \href{mailto:hpurohit@gmu.edu}{hpurohit@gmu.edu}) are with George Mason University, Fairfax, VA, USA.}%
\thanks{Michin Hong (email: \href{mailto:hongmi@iu.edu}{hongmi@iu.edu}) is with Indiana University, Bloomington, IN, USA.}%
}

\markboth{IEEE Transactions on Emerging Topics in Computational Intelligence,~Vol.~xx, No.~x, x-month~2026}{} %


\maketitle

\begin{abstract}
The rapid growth of social media participation has increased exposure to socially engineered cyber threats (e.g., phishing, romance fraud, tech-support scams), yet prevailing assessment tools remain fragmented: the \emph{Common Vulnerability Scoring System (CVSS)} is primarily technical and largely omits human susceptibility, while the \emph{Social Vulnerability Index (SVI)} is community-oriented and lacks cyber-specific modeling. To address emerging needs in computational intelligence for socio-cyber risk assessment, this paper proposes the \emph{Social Cyber Vulnerability Index (SCVI)}, an interpretable, uncertainty-aware metric that fuses two complementary components: (i) an \emph{Individual Vulnerability Index (IVI)} capturing awareness/knowledge, behavioral patterns, psychological factors, and prior victimization experience, and (ii) an \emph{Attack Severity Index (ASI)} capturing attack frequency, consequences, and sophistication. We instantiate and validate SCVI across heterogeneous modalities: a nationally scoped survey (iPoll; 4,596 U.S. adults) and social-media narratives (450 Reddit \texttt{r/scams} reports, 2016--2024), showing SCVI can be computed from both structured questionnaires and CI-driven feature extraction from text (linguistic/annotation-derived factors). Robustness is quantified via sensitivity analysis and 10,000-iteration Monte Carlo simulations, demonstrating stable rankings under plausible weight variability and revealing context-dependent drivers (e.g., experience and sophistication in iPoll versus frequency in Reddit). Comparative evaluation shows SCVI captures distinct socio-technical signals (Spearman correlation with CVSS $\rho\!=\!0.33$; with SVI $\rho\!\approx\!-0.01$) and surfaces demographic and regional disparities. \emph{Key finding:} SCVI provides substantially stronger separation between victim and non-victim groups than CVSS and SVI, enabling more reliable identification of high-risk populations and prioritization of interventions against emerging AI-enabled scams.
\end{abstract}


\section{Introduction}

\subsection{Motivation and Goal}
Social media and other online platforms are now central to communication, commerce, and community building, but they also serve as a primary venue for \emph{social cyber threats} such as phishing, impersonation, romance scams, and investment fraud. By exploiting trust and attention, these attacks cause financial loss, psychological distress, reputational harm, and erode confidence in digital ecosystems. The urgency is increasing as generative AI enables more realistic, personalized, and scalable social engineering, amplifying exposure and impact.

Yet, current risk assessment frameworks remain incomplete for this setting. The \textit{Common Vulnerability Scoring System} (CVSS)~\cite{black2008cyber} prioritizes remediation of \emph{technical} vulnerabilities, while the \textit{Social Vulnerability Index} (SVI)~\cite{flanagan2011social} captures community-level socio-economic fragilities but does not model cyber threat exposure or individual susceptibility. Consequently, stakeholders lack a dedicated, interpretable metric for quantifying \emph{individual-level} socio-cyber vulnerability where human, behavioral, and threat factors interact.

This work addresses this gap by introducing the \textit{Social Cyber Vulnerability Index} (SCVI), which bridges technical risk assessment and human-centered vulnerability by synthesizing individual susceptibility signals, such as awareness, behavioral and psychological traits, and prior experiences, with threat-level attributes such as attack frequency, sophistication, and consequence~\cite{koning2023risk,frauenstein2020susceptibility,bolpagni2022cyber}. Such a metric enables evidence-based prioritization and targeted interventions for at-risk populations.

Quantifying socio-cyber vulnerability is challenging due to limited ground truth, including underreporting and subjective outcomes, strong context dependence, and reliance on proxy indicators from heterogeneous data sources~\cite{jia2024measurement,tate2013uncertainty,schmidtlein2008sensitivity}. Prior vulnerability formulations often lack validation, adaptability, or scalability across diverse threats and populations~\cite{alim2011axioms}. Moreover, attacker tactics evolve rapidly, requiring metrics that remain robust and interpretable over time.

\textbf{The primary goal of this research} is to develop and validate SCVI as a scalable framework that integrates individual-level susceptibility with threat-level attributes using complementary evidence from survey data (iPoll) and social media interactions (Reddit)~\cite{nguyen2021analysis,g2024spatiotemporal}. SCVI supports policymakers, practitioners, and platforms by enabling comparable vulnerability estimates for adaptive cybersecurity strategies and targeted allocation of protective resources.

A preliminary version of this work appeared in our 4-page workshop paper~\cite{mitra24-workshop}. The present manuscript substantially extends that publication in both scope and technical depth. Specifically, we (i) introduce the full Social Cyber Vulnerability Index (SCVI) that explicitly fuses an Individual Vulnerability Index (IVI) with an Attack Severity Index (ASI), whereas the workshop version focused primarily on user-oriented vulnerability dimensions; (ii) operationalize the framework on heterogeneous modalities, combining a nationally scoped iPoll survey with Reddit \texttt{r/scams} narratives via computational feature extraction, rather than a survey-only demonstration; and (iii) provide significantly expanded evaluation, including robustness and uncertainty analyses (sensitivity and 10{,}000-iteration Monte Carlo), benchmarking against SVI/CVSS, and demographic and geographic disparity analyses. These additions yield new empirical insights and demonstrate the generalizability and practical utility of SCVI for socio--cyber risk assessment and targeted intervention design.

\subsection{Key Contributions}
This work makes the following key contributions:
\begin{enumerate}
    \item We propose the \textit{Social Cyber Vulnerability Index} (SCVI), a unified metric that integrates individual-level susceptibility (e.g., awareness/knowledge, behavioral and psychological factors, past experience) with attack-level characteristics (e.g., frequency, consequence, sophistication), addressing gaps not captured by CVSS~\cite{black2008cyber} or SVI~\cite{flanagan2011social}.
    
    \item We demonstrate SCVI’s applicability across heterogeneous data modalities by operationalizing and evaluating it using both iPoll survey data~\cite{ipoll-dataset2020} and Reddit scam report data~\cite{ftc_2020}, enabling more granular characterization of vulnerability patterns across populations and contexts.

   \item We conduct robustness and comparative analyses, including sensitivity, weight-variability, and Monte Carlo simulations, to quantify how individual and attack-level factors shape SCVI and to benchmark SCVI against CVSS and SVI, showing improved socio-cyber coverage while remaining consistent with technical-risk signals.

    \item We derive actionable implications for more inclusive cybersecurity practice by using SCVI to identify high-risk groups and scam typologies (e.g., phishing, financial fraud, romance scams, shopping fraud, tech support scams), and we outline extensions to broaden coverage, improve weighting via data-driven calibration, and adapt to emerging AI-enabled threats.

    \item We report empirical findings on SCVI behavior, including which susceptibility and threat attributes most strongly drive vulnerability and how SCVI responds under weight perturbations and uncertainty, yielding interpretable evidence for prioritizing interventions and refining socio-cyber risk assessment.

\end{enumerate}

\begin{table*}[t]
\centering
\caption{Comparison of representative vulnerability indices and the proposed SCVI.}
\label{tab:cvss_svi_scvi}
\begin{tabular}{lccc}
\hline
\textbf{Aspect} & \textbf{CVSS}~\cite{black2008cyber} & \textbf{SVI}~\cite{flanagan2011social} & \textbf{SCVI (this work)} \\
\hline
Primary purpose & Technical vulnerability severity & Community social vulnerability & Individual socio-cyber vulnerability \\
Unit of analysis & Software/system asset & Geographic community/region & Individual (with context) \\
Threat focus & Exploitable software flaws & Broad societal stressors & Social cyber threats and scams \\
Key inputs & Exploitability, CIA impacts & Socio-economic indicators & Susceptibility factors + threat attributes \\
Accounts for attacker tactics & Limited & No & Yes, via threat attributes \\
Data modality & Technical scoring & Census and community data & Survey + social media evidence \\
Typical output use & Patch prioritization & Resource planning & Targeted interventions and prioritization \\
\hline
\end{tabular}%
\end{table*}
\section{Related Work}\label{sec:related-work}
This section reviews prior work on vulnerability metrics, individual-level predictors of susceptibility, and spatiotemporal scam and fraud analyses to contextualize the proposed SCVI.

\subsection{Socio-Technical and Behavioral Vulnerability Metrics}
\label{subsec:vul-metrics}
\citet{black2008cyber} evaluated CVSS as a standardized approach for assessing software vulnerabilities based on exploitability and impacts on confidentiality, integrity, and availability. In contrast, \citet{flanagan2011social} introduced the Social Vulnerability Index (SVI) to identify vulnerable communities using socio-economic indicators, but it is not designed to represent cyber threat exposure or susceptibility. Bridging socio-technical aspects, \citet{bolpagni2022cyber} proposed a Cyber Risk Index linking national development indicators with cyber risk, yet it does not capture social-media-centric manipulation and targeting. At the individual level, \citet{frauenstein2020susceptibility} examined personality correlates of phishing susceptibility, but did not account for broader contextual and cultural factors that shape risk.  More broadly, \citet{bhol2023taxonomy} organized cybersecurity metrics across vulnerabilities, protections, threats, users, and situations, and used Multi-Criteria Decision Making (MCDM) for prioritization. Similarly, \citet{van2021respite} reviewed socio-technical metrics for SMEs and introduced SYMBALS to support prioritization and evaluation. For social engineering, \citet{alim2011axioms} proposed \textit{Individual Vulnerability} (VI) and \textit{Relative Vulnerability} (VR), but these formulations require empirical validation and may oversimplify heterogeneous real-world settings.

Overall, existing metrics either emphasize technical remediation (e.g., CVSS) or community-level fragility (e.g., SVI), while individual susceptibility and attack attributes are typically treated in isolation. These limitations motivate an index that can quantify individual-level socio-cyber vulnerability using interpretable constructs and heterogeneous evidence.

Table~\ref{tab:cvss_svi_scvi} summarizes key differences among CVSS, SVI, and the proposed SCVI in terms of purpose, unit of analysis, input factors, and intended use, highlighting why a dedicated individual-level socio-cyber vulnerability metric is needed.

\subsection{Key Predictors of Vulnerability}
Prior studies have examined predictors of fraud and scam susceptibility across demographics, cognition, and personality. Using Dutch survey data, \citet{koning2023risk} analyzed socio-demographics, personality traits, and internet activity, reporting higher victimization risk for frequent internet users and individuals with low self-control. Focusing on older adults, \citet{sur2021contextual} found that well-being and cognitive ability were associated with lower fraud risk, whereas negative life events and loneliness increased risk. \citet{williams2023demographic} linked credit card fraud vulnerability to financial confidence, training, marital status, and homeownership. \citet{judges2017role} further highlighted the role of cognitive ability and honesty-humility in susceptibility among older adults.

These studies are often constrained by post-incident measurement~\cite{koning2023risk}, self-reported outcomes~\cite{sur2021contextual,williams2023demographic}, and limited sample diversity~\cite{judges2017role}. Moreover, while they identify correlates of susceptibility, they typically do not produce a unified vulnerability score that is comparable across populations and can be integrated with threat-level attributes. This gap motivates an index-based approach that translates predictors into actionable, interpretable vulnerability estimates.

\subsection{Geospatial and Temporal Scam Patterns}

Several studies have examined how scam and fraud patterns vary across geography and time. \citet{edwards2018geography} analyzed online dating scam profiles to identify major sources and reuse patterns using geolocation-linked signals. \citet{nguyen2021analysis} used regression to relate regional crime patterns to demographic disparities. \citet{g2024spatiotemporal} analyzed U.S. financial fraud trends (2018--2022), reporting spatial clustering and post-COVID shifts using statistics such as Moran's I.

However, these approaches commonly face dataset bias and proxy limitations~\cite{edwards2018geography}, reliance on secondary data without complementary individual-level evidence~\cite{nguyen2021analysis}, and underreporting that can distort spatial inference~\cite{g2024spatiotemporal}. While they illuminate exposure and macro-level patterns, they do not quantify individual socio-cyber vulnerability or integrate susceptibility with attack attributes in a unified metric.

\vspace{1mm}
\noindent \textbf{Summary and positioning.}
Motivated by these gaps, the proposed SCVI integrates individual-level susceptibility (behavioral, psychological, and demographic factors) with threat-level attributes (e.g., frequency and sophistication), and operationalizes vulnerability using heterogeneous evidence from survey and social media data. This design supports context-aware, interpretable, and comparable socio-cyber vulnerability assessment across populations and settings.

\begin{SCfigure*}[0.35][t]
\centering
\resizebox{0.72\textwidth}{!}{%
\begin{tikzpicture}[
  font=\sffamily,
  >=Latex,
  node distance=7mm and 12mm,
  box/.style={
    rounded corners=3pt,
    draw=black!70,
    very thick,
    align=center,
    inner xsep=10pt,
    inner ysep=8pt,
    blur shadow={shadow xshift=0.6pt, shadow yshift=-0.6pt, shadow blur steps=4}
  },
  pill/.style={
    rounded corners=8pt,
    draw=black!70,
    very thick,
    fill=yellow!35,
    align=center,
    inner xsep=10pt,
    inner ysep=6pt,
    blur shadow={shadow xshift=0.6pt, shadow yshift=-0.6pt, shadow blur steps=4}
  },
  split/.style={
    rounded corners=3pt,
    draw=black!70,
    very thick,
    align=left,
    inner xsep=10pt,
    inner ysep=8pt,
    minimum width=3cm,
    blur shadow={shadow xshift=0.6pt, shadow yshift=-0.6pt, shadow blur steps=4}
  },
  arr/.style={->, very thick, draw=black!75}
]

\node[pill] (lit) {\textbf{Evidence Synthesis}\\[-1pt]\footnotesize Literature review \& threat landscape};

\node[box, fill=cyan!20, below=0.7cm of lit, text width = 5cm] (pred) {\textbf{Key Predictors of Social--Cyber Vulnerability}\\[-1pt]\footnotesize
Identify measurable drivers that explain \emph{who} is at risk and \emph{how} attacks cause harm};

\node[split, fill=orange!20, below left=1cm of pred, align=center, text width = 4cm] (ivi) {%
  \textbf{Individual\\  Susceptibility Index (IVI)}\\\scriptsize
 
    Awareness and knowledge $\bullet$ 
    Behavioral \& psychological traits $\bullet$ 
  Prior experience with cyber scams
  
};

\node[split, fill=orange!10, below right=of pred, align=center, text width = 3.5cm] (asi) {%
  \textbf{Attack \\ Severity \\ Index (ASI)}\\\scriptsize
 Frequency 
    $\bullet$ Consequence \& impact
    $\bullet$ Sophistication
};

\node[box, fill=teal!18, below=13mm of pred, text width=4cm] (design) {\textbf{SCVI Formulation}\\[-1pt]\footnotesize
Fuse IVI and ASI into a unified index \\ (normalization, weighting, aggregation)};

\node[box, fill=gray!12, below=of design] (est) {\textbf{Index Instantiation from Data}\\[-1pt]\footnotesize
Survey (iPoll) and social media reports (Reddit \texttt{r/scams}) to estimate IVI/ASI factors};

\node[pill, below=of est] (score) {\textbf{SCVI Scores}\\[-1pt]\footnotesize Individual- and group-level vulnerability estimates};

\node[box, fill=green!14, below=of score, text width = 7cm] (eval) {\textbf{Robustness, Validation, and Insights}\\\scriptsize
Robustness \& sensitivity testing  \\ \scriptsize Comparison vs. SVI/CVSS  \\[-4pt]\scriptsize  Disparities \& actionable prioritization};

\draw[arr] (lit) -- (pred);
\draw[arr] (pred) -- (ivi);
\draw[arr] (pred) -- (asi);
\draw[arr] (ivi.east) --  (design.west);
\draw[arr] (asi.west) --  (design.east);
\draw[arr] (design) -- (est);
\draw[arr] (est) -- (score);
\draw[arr] (score) -- (eval);

\end{tikzpicture}
}
\caption{\textbf{Overview of the SCVI development pipeline.} Evidence synthesis from prior literature and the social-cyber threat landscape identifies predictors of vulnerability, organized into an \emph{Individual Susceptibility Index (IVI)} (e.g., awareness/knowledge, behavioral traits, prior scam exposure) and an \emph{Attack Severity Index (ASI)} (e.g., frequency, impact, sophistication). IVI and ASI are normalized, weighted, and aggregated to form the \emph{Social Cyber Vulnerability Index (SCVI)}, which is instantiated from survey and social media scam reports and evaluated via robustness and sensitivity analyses to surface disparities and prioritize interventions against emerging AI-enabled scams.
}

\label{fig:scvi_overview}
\end{SCfigure*}
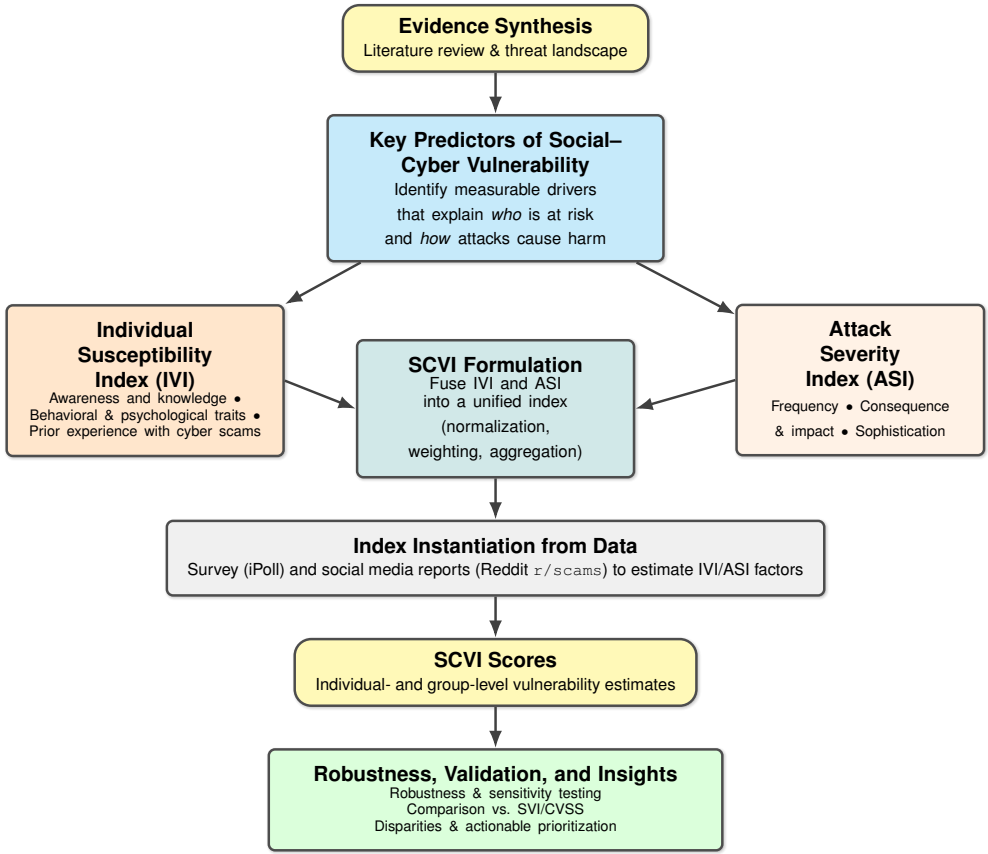

\section{Proposed Social Cyber Vulnerability Index}\label{sec:proposed}

Fig.~\ref{fig:scvi_overview} summarizes the SCVI development pipeline. We first review prior literature to identify predictors of socio-cyber vulnerability, and organize them into \emph{Individual Vulnerability} (e.g., awareness and knowledge, behavioral and psychological patterns, prior cyber-scam experience) and \emph{Attack Severity} (e.g., attack frequency, impact, sophistication). These factors guide SCVI construction, which we estimate from complementary evidence sources including survey responses and social media data. We evaluate SCVI via robustness and sensitivity analyses and comparative result analysis to assess how effectively it differentiates victim risk across populations.

\subsection{Considered Key Vulnerability Factors to Social Scams}
\label{subsec:key-vul-factors-sc}

Social scams are a specialized form of online fraud~\cite{kavrestad2018defining} that targets victims through interactive channels such as email, websites, social media, and chat rooms. Unlike fraud that exploits technical weaknesses, social scams exploit trust, social dynamics, and persuasion to bypass behavioral and psychological defenses~\cite{guo2020online,susser2019online}. Risk therefore depends on both user susceptibility and attack severity.

To capture this socio-technical interplay, we structure SCVI around two complementary components: an \textit{Individual Vulnerability Index} (IVI) that represents user susceptibility, and an \textit{Attack Severity Index} (ASI) that represents the severity of scam exposure. Together, the seven predictors used to construct IVI and ASI are referred to as the \textit{ABPE--FCS predictor set} (IVI: ABPE; ASI: FCS), and they form the factor basis of the proposed \textit{Social Cyber Vulnerability Index} (SCVI).

\paragraph{Individual Vulnerability Index (IVI).}
IVI summarizes individual susceptibility using four predictors:
\begin{enumerate}
\item \textbf{Individual Awareness and Knowledge (A):} The extent to which individuals understand social cyber threats and available protective measures. Limited knowledge of scam tactics and defenses increases susceptibility~\cite{albladi2020predicting}.
\item \textbf{Behavioral Patterns (B):} An individual’s online activities and day-to-day security practices. Greater exposure to high-risk contexts and weaker protective behaviors raise victimization likelihood~\cite{reyns2013security}.

\item \textbf{Psychological Factors (P):} Traits and perceptions such as trust propensity and risk perception. Excessive trust in unfamiliar communications or attenuated risk perception increases susceptibility~\cite{Robb2023}.
\item \textbf{Past Experience (E):} Prior encounters with scams and the resulting coping or recovery strategies. Experience may build resilience, but inadequate recovery can leave persistent vulnerabilities~\cite{fredrick2021resiliency}.
\end{enumerate}

\paragraph{Attack Severity Index (ASI).}
ASI characterizes attack-side conditions that modulate risk using three predictors:
\begin{enumerate}
\item \textbf{Frequency of Attacks (F):} The number of scam attempts a user encounters. Higher frequency increases the chance of at least one successful attempt and may reduce vigilance through repeated exposure~\cite{whitty2020there}.
\item \textbf{Consequences of Attacks (C):} The magnitude of harm (e.g., financial or emotional), which can amplify downstream vulnerability and lasting impact~\cite{norris2019psychology}.
\item \textbf{Sophistication of Attacks (S):} The realism and personalization of scam content, including mimicry of legitimate entities and use of personal information. More sophisticated attacks are harder to detect and increase susceptibility~\cite{collier2022sophisticated}.
\end{enumerate}

By integrating the ABPE--FCS predictor set from user- and attack-side perspectives, SCVI provides a holistic measure of susceptibility to social cyberattacks. While many factors may influence social scam risk, this study focuses on seven predictors most consistently supported in prior work, and operationalizes them using survey and social media datasets.

\subsection{Social Cyber Vulnerability Index (SCVI)}
We propose the \emph{Social Cyber Vulnerability Index} (SCVI) to quantify an individual’s susceptibility to social cyber scams by jointly modeling (i) user-side susceptibility and (ii) attack-side severity. For an individual $i$ and scam/attack type $k$, SCVI is defined as a convex combination of the \emph{Individual Vulnerability Index} (IVI) and the \emph{Attack Severity Index} (ASI):
\begin{equation}
\label{eq:scvi}
\mathrm{SCVI}_{i,k} = \alpha\,\mathrm{IVI}_{i,k} + \beta\,\mathrm{ASI}_{i,k},
\end{equation}
where $\alpha,\beta \in [0,1]$ and $\alpha+\beta=1$. Here, $\mathrm{IVI}_{i,k}$ captures how susceptible individual $i$ is to scam $k$, while $\mathrm{ASI}_{i,k}$ captures how severe scam $k$ is as perceived by individual $i$.

\paragraph{Individual Vulnerability Index (IVI)}
IVI aggregates four predictors (ABPE) as:
\begin{equation}
\label{eq:ivi}
\mathrm{IVI}_{i,k} = w_{A_{i,k}}\,\mathrm{A}_{i,k} + w_{B_{i,k}}\,\mathrm{B}_{i,k} + w_{P_{i,k}}\,\mathrm{P}_{i,k} + w_{E_{i,k}}\,\mathrm{E}_{i,k},
\end{equation}
where $w_{A_{i,k}} + w_{B_{i,k}} + w_{P_{i,k}} + w_{E_{i,k}} = 1$. Each factor is decomposed into two subcomponents:
\begin{eqnarray}
\label{eq:ivi_components}
\mathrm{A}_{i,k} = \mathrm{A}^{A}_{i,k} + \mathrm{A}^{K}_{i,k}, \quad
\mathrm{B}_{i,k} = \mathrm{B}^{R}_{i,k} + \mathrm{B}^{S}_{i,k}, \\
\mathrm{P}_{i,k} = \mathrm{P}^{C}_{i,k} + \mathrm{P}^{I}_{i,k}, \quad
\mathrm{E}_{i,k} = \mathrm{E}^{E}_{i,k} + \mathrm{E}^{R}_{i,k}.  \nonumber
\end{eqnarray}
Specifically, $\mathrm{A}_{i,k}$ represents (lack of) awareness and knowledge, combining unfamiliarity with scam $k$ ($\mathrm{A}^{A}_{i,k}$) and knowledge of protective measures ($\mathrm{A}^{K}_{i,k}$). $\mathrm{B}_{i,k}$ represents behavioral patterns, combining the frequency of risk-enhancing behaviors ($\mathrm{B}^{R}_{i,k}$) and protective security practices ($\mathrm{B}^{S}_{i,k}$). $\mathrm{P}_{i,k}$ represents psychological factors, combining trust propensity in scam-related communications ($\mathrm{P}^{C}_{i,k}$) and risk perception or impulsivity ($\mathrm{P}^{I}_{i,k}$). $\mathrm{E}_{i,k}$ represents past experience, combining prior encounters with scam $k$ ($\mathrm{E}^{E}_{i,k}$) and the individual’s response or recovery strategies ($\mathrm{E}^{R}_{i,k}$).

\paragraph{Attack Severity Index (ASI)}
ASI aggregates three predictors (FCS) as:
\begin{equation}
\label{eq:asi}
\mathrm{ASI}_{i,k} = w_{F_{i,k}}\,\mathrm{F}_{i,k} + w_{C_{i,k}}\,\mathrm{C}_{i,k} + w_{S_{i,k}}\,\mathrm{S}_{i,k},
\end{equation}
where $w_{F_{i,k}} + w_{C_{i,k}} + w_{S_{i,k}} = 1$. We further define:
\begin{equation}
\label{eq:asi_components}
\begin{aligned}
\mathrm{F}_{i,k} &= \mathrm{F}^{TA}_{i,k} + \mathrm{F}^{AA}_{i,k}, \quad
\mathrm{C}_{i,k} = \mathrm{C}^{FI}_{i,k} + \mathrm{C}^{PI}_{i,k} + \mathrm{C}^{SI}_{i,k}, \\
\mathrm{S}_{i,k} &= \mathrm{S}^{C}_{i,k} + \mathrm{S}^{SE}_{i,k}.
\end{aligned}
\end{equation}
Here, $\mathrm{F}_{i,k}$ captures attack frequency via attempted attacks ($\mathrm{F}^{TA}_{i,k}$) and attacks actually encountered ($\mathrm{F}^{AA}_{i,k}$). $\mathrm{C}_{i,k}$ captures consequences of a successful attack through financial impact ($\mathrm{C}^{FI}_{i,k}$), psychological or emotional impact ($\mathrm{C}^{PI}_{i,k}$), and personal-safety impact ($\mathrm{C}^{SI}_{i,k}$). $\mathrm{S}_{i,k}$ captures perceived sophistication via mimicry of legitimate communications ($\mathrm{S}^{C}_{i,k}$) and the use of personalization or advanced social engineering ($\mathrm{S}^{SE}_{i,k}$).

\paragraph{Scaling and operationalization}
We represent each component (e.g., $\mathrm{A}_{i,k}$) on an integer scale in $[0,5]$. In practice, $\mathrm{IVI}_{i,k}$ can be obtained from Likert-scale survey instruments or inferred from behavioral and lexical traces on social media, whereas $\mathrm{ASI}_{i,k}$ is derived from attack characteristics as perceived and reported by users~\cite{zong2019analyzing}. The weights $\alpha,\beta$ and $\{w_{\cdot}\}$ balance user-side susceptibility and attack-side severity, enabling SCVI to support comparative assessment and targeted interventions across populations and scam types.

\section{SCVI Estimation Using iPoll and Reddit Data}

This section describes how we estimate SCVI using two complementary data sources: iPoll survey responses and Reddit scam reports. iPoll provides structured survey-based measurements of user susceptibility (e.g., risky behaviors and protective knowledge), while Reddit contributes contextual, user-generated evidence about scam experiences and characteristics. By integrating these modalities, SCVI combines statistical grounding with real-world context, yielding a comprehensive and adaptable measure of social cyber vulnerability.

\subsection{SCVI Using the iPoll Dataset}

To compute the Social Cyber Vulnerability Index (SCVI), we use survey data from the iPoll dataset~\cite{ipoll-dataset2020}. The dataset, \emph{The Impostors: Stealing Money, Damaging Lives. An AARP National Survey of Adults 18+} (iPoll), examines scam awareness, experiences, and behaviors among 4{,}596 adults in the United States. The survey was administered by NORC at the University of Chicago between January~2 and January~16, 2020, using computer-assisted telephone interviews (CATI) and web-based instruments. iPoll captures diverse indicators of online behavior, personal traits, and scam exposure (e.g., romance scams, government impostor scams, and identity theft) across 113 variables, and includes weighting factors to support nationally and regionally representative analyses. These characteristics make iPoll well-suited for estimating SCVI and analyzing how vulnerability varies across demographic groups.

\subsubsection{Modeling the Individual Vulnerability Index (IVI)}
We model IVI by encoding and aggregating iPoll responses into seven subcomponents: lack of awareness ($\mathrm{A}_{i,k}^A$), lack of knowledge of protective measures ($\mathrm{A}_{i,k}^K$), frequency of risk-enhancing behaviors ($\mathrm{B}_{i,k}^R$), trust level ($\mathrm{P}_{i,k}^C$), risk perception and impulsivity ($\mathrm{P}_{i,k}^I$), past encounters ($\mathrm{E}_{i,k}^E$), and responses to past incidents ($\mathrm{E}_{i,k}^R$). Each subcomponent is computed by mapping categorical responses to ordinal scores using predefined encodings, where larger values indicate higher vulnerability. For example, for $\mathrm{A}_{i,k}^A$, a response of ``Very concerned'' is assigned 0, whereas ``Not at all concerned'' is assigned 3. For $\mathrm{A}_{i,k}^K$, incorrect protective knowledge (e.g., responding ``True'' to a false security statement) is assigned a higher score (e.g., 5), reflecting increased vulnerability.

For each subcomponent, we average the encoded scores across the associated survey items. In particular, $\mathrm{B}_{i,k}^R$ captures exposure via the frequency of online activities; $\mathrm{P}_{i,k}^C$ and $\mathrm{P}_{i,k}^I$ are derived from self-descriptions of interpersonal trust and decision-making tendencies; and $\mathrm{E}_{i,k}^E$ and $\mathrm{E}_{i,k}^R$ reflect prior scam encounters and their outcomes (e.g., financial loss and emotional distress). The final IVI is computed as the mean of the seven subcomponents:
\begin{equation}
\label{eq:ipoll_ivi}
\mathrm{IVI}_{i,k} = \frac{1}{7}\Big(\mathrm{A}_{i,k}^A + \mathrm{A}_{i,k}^K + \mathrm{B}_{i,k}^R + \mathrm{P}_{i,k}^C + \mathrm{P}_{i,k}^I + \mathrm{E}_{i,k}^E + \mathrm{E}_{i,k}^R\Big).
\end{equation}

\subsubsection{Modeling the Attack Severity Index (ASI)}
We compute ASI from iPoll by encoding and aggregating responses into three components: frequency ($\mathrm{F}_{i,k}$), consequence ($\mathrm{C}_{i,k}$), and sophistication ($\mathrm{S}_{i,k}$), where larger values indicate greater attack severity. Frequency aggregates items describing scam exposure prevalence, consequence aggregates reported financial and well-being impacts, and sophistication captures perceived plausibility and realism of scam attempts. We then compute ASI as the average of these three components:
\begin{equation}
\label{eq:ipoll_asi}
\mathrm{ASI}_{i,k} = \frac{1}{3}\Big(\mathrm{F}_{i,k} + \mathrm{C}_{i,k} + \mathrm{S}_{i,k}\Big).
\end{equation}

The detailed feature extraction and response-to-score mappings for IVI and ASI are provided in Tables~I and~II in the supplement document, respectively.

\subsection{SCVI Using the Reddit Scam Reports}

The Reddit Scam Reports dataset~\cite{ftc_2020} contains user-generated posts from the subreddit \texttt{r/scams} describing scam encounters and outcomes. It spans 2016--2023 (collected via the Pushshift API) and includes additional records from October--November 2024. From an initial pool of 5{,}000 posts, we selected 450 scam reports with balanced representation across years to support comparative temporal analysis.

We preprocess the text by replacing common slang, removing URLs, and stripping user mentions and hashtags. We further normalize posts by converting to lowercase, expanding contractions, reducing elongated words, and removing non-alphanumeric characters while preserving punctuation. The resulting corpus supports consistent annotation of scam types and outcomes and provides a clean foundation for extracting IVI- and ASI-related signals.

\subsubsection{Modeling the Individual Vulnerability Index (IVI)}
For Reddit, IVI is inferred from user-generated language and interaction traces reflecting awareness, behavior, and experience. We operationalize IVI through four components:

\begin{itemize}[leftmargin=*]
    \item \textbf{Lack of Awareness and Knowledge ($\mathrm{A}$):} We approximate awareness using annotations of scam type and reported success, together with indicators of exposure to prevention knowledge in community discussions. Participation in online communities can increase knowledge sharing and familiarity with scam tactics, which may reduce susceptibility~\cite{ayachi2021virtual}. Users who frequently engage with scam discussions may exhibit higher familiarity with scam patterns and mitigation strategies, reflecting lower vulnerability.

    \item \textbf{Behavioral Traits ($\mathrm{B}$):} We model behavioral vulnerability using interaction signals (e.g., posting behavior) and linguistic markers associated with risk-taking or impulsivity~\cite{herman2018risk}. We extract features using Linguistic Inquiry and Word Count (LIWC)~\cite{liwc22}, including indicators such as clout, perceptual language, and informal markers~\cite{tausczik2010psychological}. For example, low clout may reflect lower confidence, while heavier reliance on sensory descriptions can increase susceptibility to scams that exploit credibility cues. Consistent with prior findings, low self-confidence and over-reliance on perceptual processing can elevate vulnerability~\cite{norris2019psychology}. We also incorporate skepticism signals, such as negations (e.g., ``no,'' ``not,'' ``never''), which have been associated with resistance to misinformation and manipulative content~\cite{wright2020many}.

    \item \textbf{Psychological Factors ($\mathrm{P}$):} We capture psychological susceptibility via linguistic evidence of emotion and cognition. LIWC features such as affect and cognitive-process categories provide proxies for emotional state and analytical thinking. Stronger analytical and cognitive skills are associated with improved scam resistance~\cite{hruschka2023learning,gamble2015aging}, whereas anxiety, neuroticism, and stress can impair judgment and increase susceptibility~\cite{boyle2022degraded,cho2016effect,norris2021personality}. We further consider markers of credulity, which has been linked to fraud risk among older adults~\cite{shao2019credulity}, and treat skepticism and reflective language as protective signals.

    \item \textbf{Experience ($\mathrm{E}$):} We represent experience using annotations indicating whether the post describes a direct victimization event and whether it reports financial loss or emotional distress. Prior exposure can increase awareness and protective knowledge, but it may also reflect persistent susceptibility depending on recovery and coping responses~\cite{houtti2024survey,sheng2010falls}.
\end{itemize}

\subsubsection{Modeling the Attack Severity Index (ASI)}
We model ASI by annotating each report with two attributes: scam type and scam success. Two human annotators label the data, with additional refinement using OpenAI's API. Scam types include \textit{phishing}, \textit{investment scams}, \textit{lottery scams}, \textit{tech support scams}, \textit{romance scams}, \textit{online shopping scams}, \textit{job scams}, and \textit{undetected}. Scam success is encoded as a binary variable (0/1). We then derive the ASI components as follows:
\begin{itemize}
    \item \textbf{Frequency ($\mathrm{F}$):} We estimate prevalence by counting the number of reports per scam type, capturing how frequently each scam category appears in the dataset~\cite{houtti2024survey}.
    \item \textbf{Consequence ($\mathrm{C}$):} We estimate impact by combining annotated financial-loss indicators~\cite{lwin2023supporting} with measures of emotional distress extracted from text (e.g., emotion recognition and aggregation of negative and positive affect)~\cite{whitty2016online,safari2023emotion}.
    \item \textbf{Sophistication ($\mathrm{S}$):} We approximate sophistication using the success rate within each scam type, defined as the proportion of successful incidents among all reports of that type. Higher success rates suggest more effective deception and greater attack sophistication~\cite{darem2024beyond}.
\end{itemize}

\section{Evaluation Setup}

This section describes the experimental setup used to evaluate SCVI, including the data sources, weighting strategies, simulation procedures, and baseline comparisons against established indices (CVSS and SVI).

\subsection{Sensitivity Analysis of Weighting Schemes on SCVI Using iPoll and Reddit Data}
We conduct a sensitivity analysis to quantify how alternative weighting schemes affect SCVI scores across iPoll and Reddit scam reports. We vary the IVI weights $(w_A,w_B,w_P,w_E)$ and ASI weights $(w_F,w_C,w_S)$ and evaluate their effects on the mean SCVI and its variability. We then visualize and quantify each weight’s contribution to SCVI variation, revealing the most influential vulnerability dimensions.

\subsection{Monte Carlo Simulation for Weight Variability and Uncertainty in SCVI Scores}
To characterize uncertainty due to weight specification, we use Monte Carlo simulations that randomly perturb the IVI and ASI weights within plausible ranges while enforcing simplex constraints (weights within each index sum to one). Over 10{,}000 iterations, we recompute SCVI and analyze the resulting distributions to assess stability, quantify variability, and identify influential weight configurations. This provides a robust view of SCVI behavior under weighting uncertainty.

\subsection{Evaluation of SCVI Against CVSS and SVI}
We benchmark SCVI against the Common Vulnerability Scoring System (CVSS)~\cite{NISTCVSSMetrics} and the Social Vulnerability Index (SVI)~\cite{CDCSVI}. CVSS rates software vulnerabilities via Base, Temporal, and Environmental metrics; we construct CVSS-like baselines by mapping iPoll fraud attributes to analogous CVSS factors (e.g., user interaction and impact). SVI measures community-level vulnerability using sociodemographic and infrastructure indicators; we derive SVI-based baselines from iPoll demographics for comparison with SCVI.

We then compute SCVI and compare it with the CVSS- and SVI-based baselines derived from iPoll, evaluating alignment and divergence across groups. Results are reported using statistical tests, correlation analyses, and outlier comparisons across key demographic categories (e.g., age, race/ethnicity, and gender), assessing the extent to which SCVI bridges cyber-behavioral and sociodemographic vulnerability signals.

\section{Analyses of Evaluation Results}

SCVI was estimated using the iPoll and Reddit datasets and compared with two existing metrics: SVI and CVSS.

\subsection{Analyses of SCVI}

\subsubsection{Analysis of the iPoll Dataset}
\begin{figure*}[t]
    \centering
    \subfloat[Distributions of IVI factors]{
        \includegraphics[width=0.3\textwidth, height=0.25\textwidth]{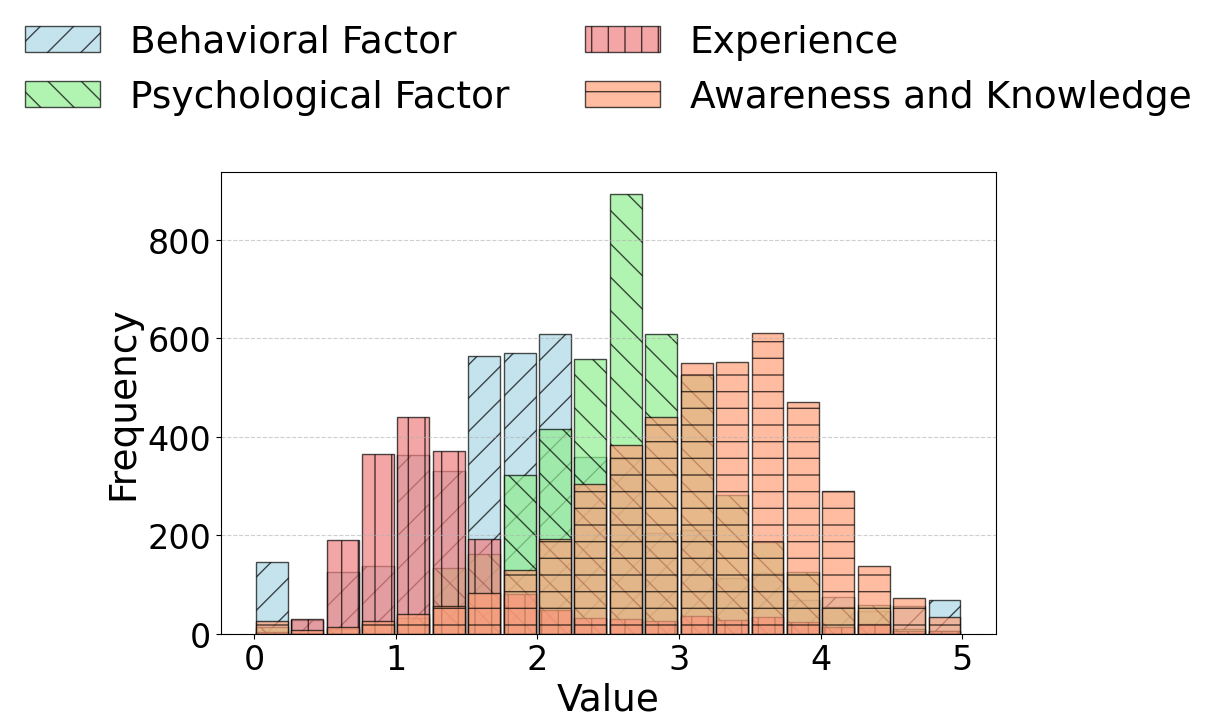}
        \label{fig:ipoll_ivi_factors}
    }
    \hspace{1mm}
    \subfloat[Distributions of IVI and ASI values]{
        \includegraphics[width=0.3\textwidth, height=0.25\textwidth]{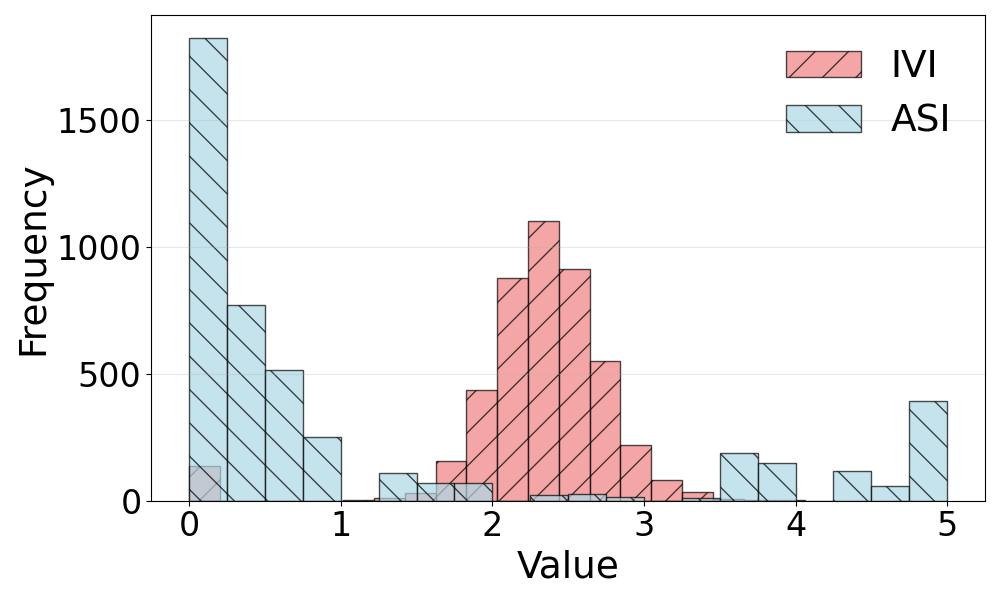}
        \label{fig:ipoll_ivi_asi}
    }
    \hspace{1mm}
    \subfloat[Distributions of ASI factors]{
        \includegraphics[width=0.3\textwidth, height=0.25\textwidth]{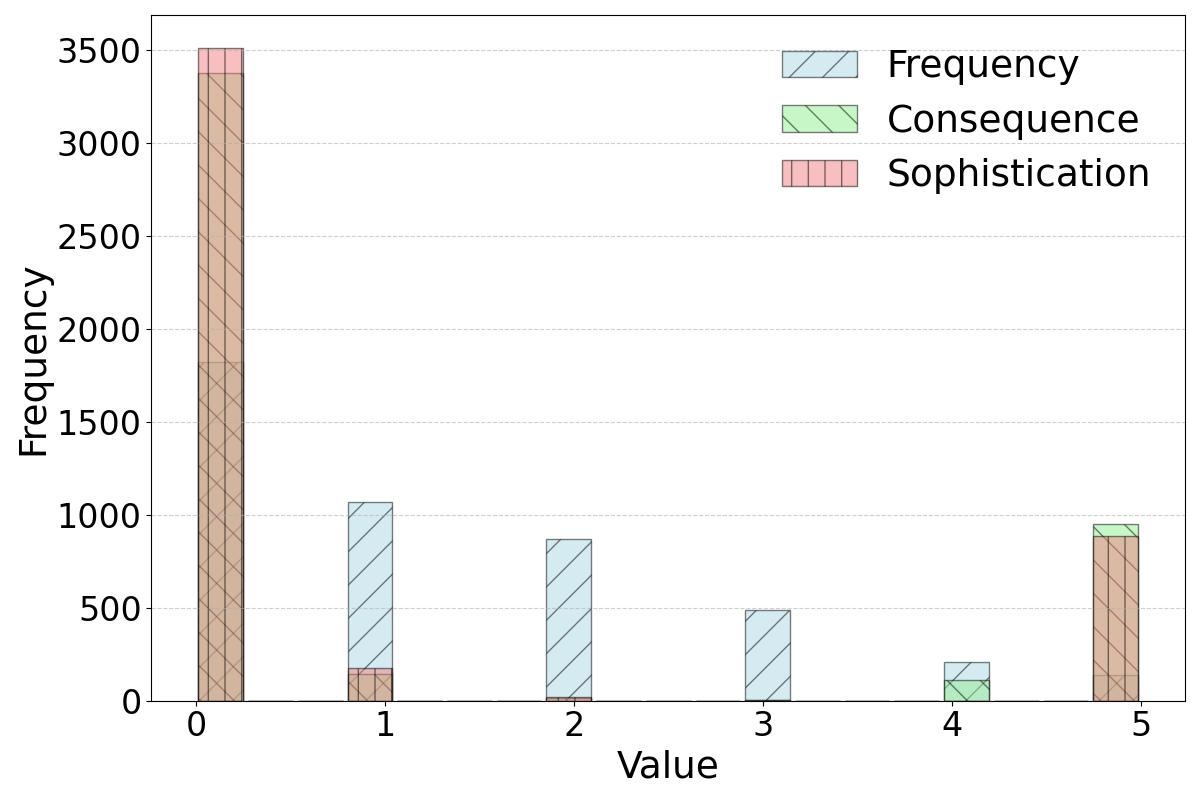}
        \label{fig:ipoll_asi_factors}
    }
    \caption{Comparison of IVI and ASI distributions and their contributing factors using the iPoll dataset. }
    \label{fig:combined-ipoll}
\end{figure*}

The analysis of the iPoll dataset reveals key patterns in user vulnerability and perceived attack severity. Fig.~\ref{fig:ipoll_ivi_factors} presents the distribution of the Individual Vulnerability Index (IVI) components. The \textit{Behavioral} factor is concentrated around a score of 1, indicating that most users report minimal engagement in risky behaviors associated with scam susceptibility. In contrast, the \textit{Psychological} factor displays a broader distribution, with the majority of users scoring between 2 and 4, suggesting moderate psychological traits linked to vulnerability.  The \textit{Experience} factor is notably skewed toward lower values, implying limited prior exposure to scam incidents among respondents. Finally, the \textit{Awareness and Knowledge} factor spans a wider range, with many users exhibiting low to moderate familiarity with scam types and warning signs, highlighting a potential gap in scam literacy.

Fig.~\ref{fig:ipoll_ivi_asi} compares the distributions of the IVI and ASI. IVI values cluster between 2 and 3, indicating moderate vulnerability for most users. In contrast, ASI is strongly skewed toward low values, with a large mass at 0, suggesting many users encounter scams with minimal severity. A smaller but notable fraction exhibits ASI near 4--5, indicating highly severe incidents. This disparity highlights the dual challenge of mitigating broad moderate vulnerability while addressing concentrated high-severity harm.

Further insights emerge from the ASI components in Fig.~\ref{fig:ipoll_asi_factors}. \textit{Frequency} is bimodal with peaks near 0 and 5, suggesting users either rarely or frequently encounter scams. \textit{Consequence} clusters around moderate values, indicating impacts are typically non-trivial but not catastrophic. \textit{Sophistication} is broadly distributed with mass near both ends, implying some scams are easily detected while others are highly deceptive.

In conclusion, the iPoll analysis indicates that moderate user vulnerability often coexists with heterogeneous, and sometimes severe, scam exposure. Even when high-risk behaviors are not pervasive, the presence of persuasive and high-impact scam attempts underscores the need for targeted awareness and mitigation efforts to reduce exposure and strengthen resilience among vulnerable populations.

\subsubsection{Analysis of Reddit Dataset}
\begin{figure*}[t]
    \centering
    \subfloat[Distributions of IVI factors]{
        \includegraphics[width=0.3\textwidth, height=0.25\textwidth]{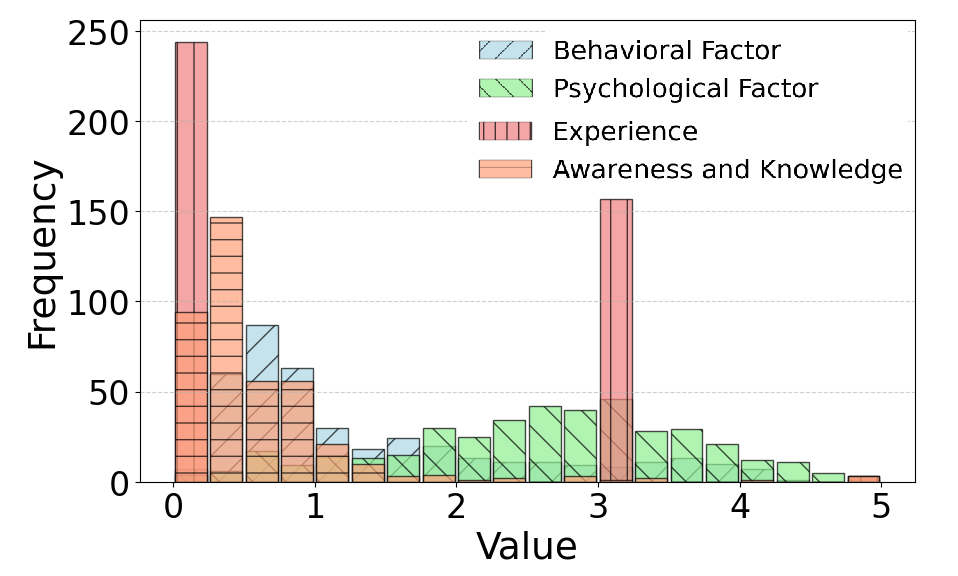}
        \label{fig:ivi_factors}
    }
    \hspace{1mm}
    \subfloat[Distributions of IVI and ASI values]{
        \includegraphics[width=0.3\textwidth, height=0.25\textwidth]{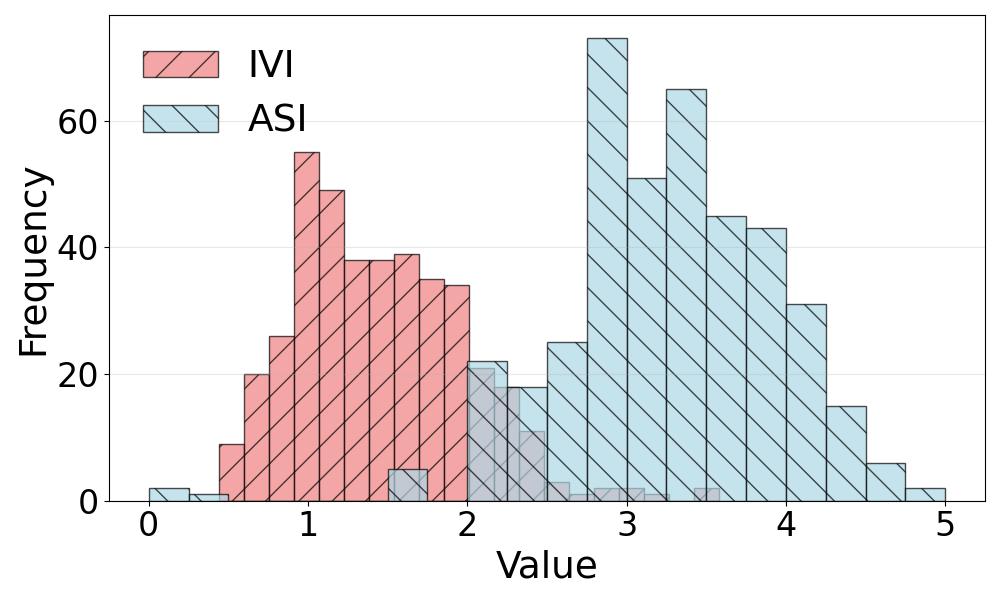}
        \label{fig:ivi_asi}
    }
    \hspace{1mm}
    \subfloat[Distributions of ASI factors]{
    \includegraphics[width=0.3\textwidth, height=0.25\textwidth]{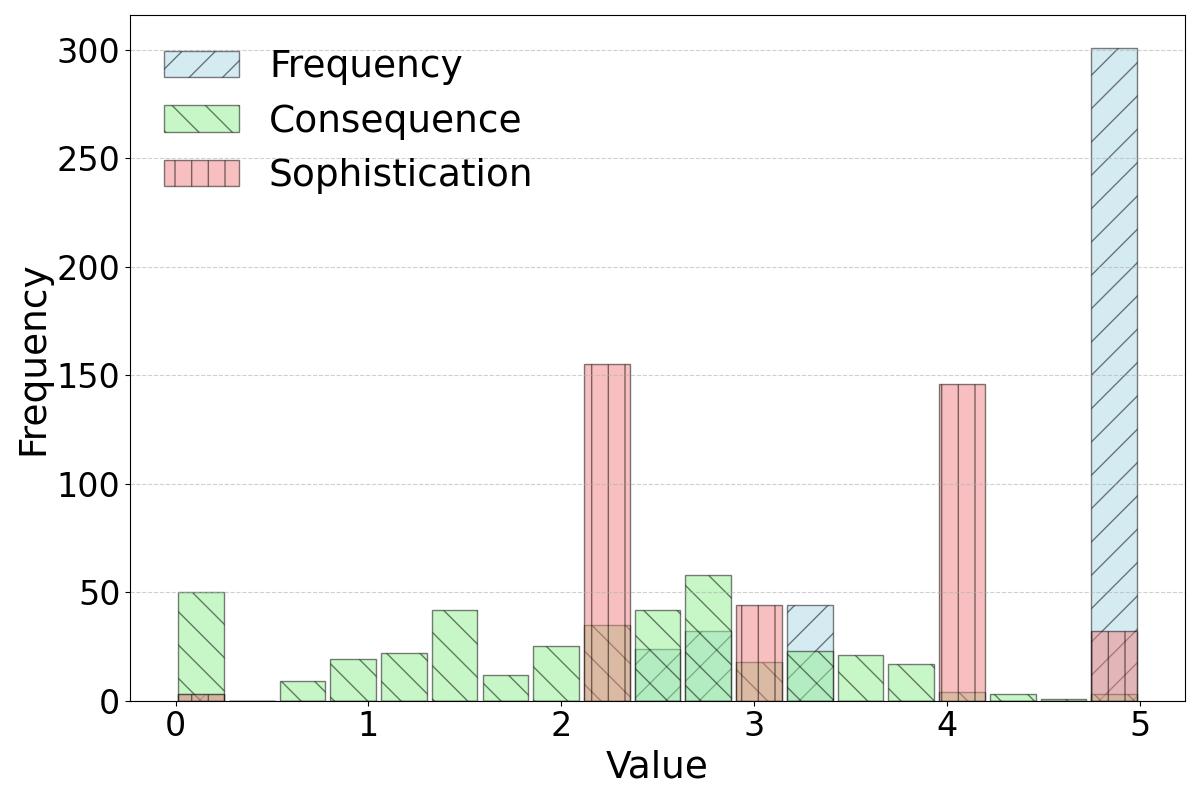}
        \label{fig:asi_factors}
    }
    \caption{Comparison of IVI and ASI distributions and their contributing factors using the Reddit dataset}
    \label{fig:combined-reddit}
\end{figure*}

The distribution of the Individual Vulnerability Index (IVI) components derived from Reddit data offers key insights into user susceptibility, as illustrated in Fig.~\ref{fig:ivi_factors}. The \textit{Behavioral} factor is strongly skewed toward lower values, indicating that most users exhibit minimal engagement in behaviors that increase scam vulnerability. In contrast, the \textit{Psychological} factor presents a more balanced distribution, with most users scoring between 2 and 4, suggesting moderate psychological influences on vulnerability.

The \textit{Experience} factor is concentrated at discrete values, with a large portion of users reporting no prior encounters with scams, while a smaller segment exhibits high scores, reflecting extensive past interactions. Finally, the \textit{Awareness and Knowledge} factor is heavily skewed toward the lower end of the scale, revealing a widespread lack of scam-related awareness and understanding among Reddit users.

Fig.~\ref{fig:ivi_asi} shows a clear contrast between IVI and ASI. IVI clusters between 1 and 2, indicating low to moderate vulnerability for most users, whereas ASI skews higher, with most scores between 3 and 4. Thus, even less vulnerable users often face substantial scam severity, motivating targeted interventions to bridge susceptibility and threat impact.

Further insights are provided by the ASI component distributions shown in Fig.~\ref{fig:asi_factors}. The \textit{Frequency} factor exhibits a highly polarized pattern, with peaks near 0 and 5, indicating that users report either very frequent or infrequent scam encounters. The \textit{Consequence} factor peaks moderately around 3, reflecting a generally moderate impact of scams across users. The \textit{Sophistication} factor displays a bimodal distribution with peaks near 2 and 4, highlighting the presence of both low-persuasion and highly convincing scam attempts.

In summary, the analysis reveals a mismatch between user vulnerability and attack severity. While most users have low IVI scores, higher ASI values indicate frequent exposure to impactful and sophisticated scams. Addressing this gap requires enhanced user education and psychological preparedness, particularly for recognizing and mitigating high-severity threats. These findings inform targeted strategies to reduce scam-related harm among vulnerable groups.

\subsection{Sensitivity Analysis of IVI and ASI Effects on SCVI}

\begin{figure*}[t]
\vspace{-5mm}
    \centering
    \subfloat[Varying the weights of the IVI factors]{
        \includegraphics[width=0.45\textwidth]{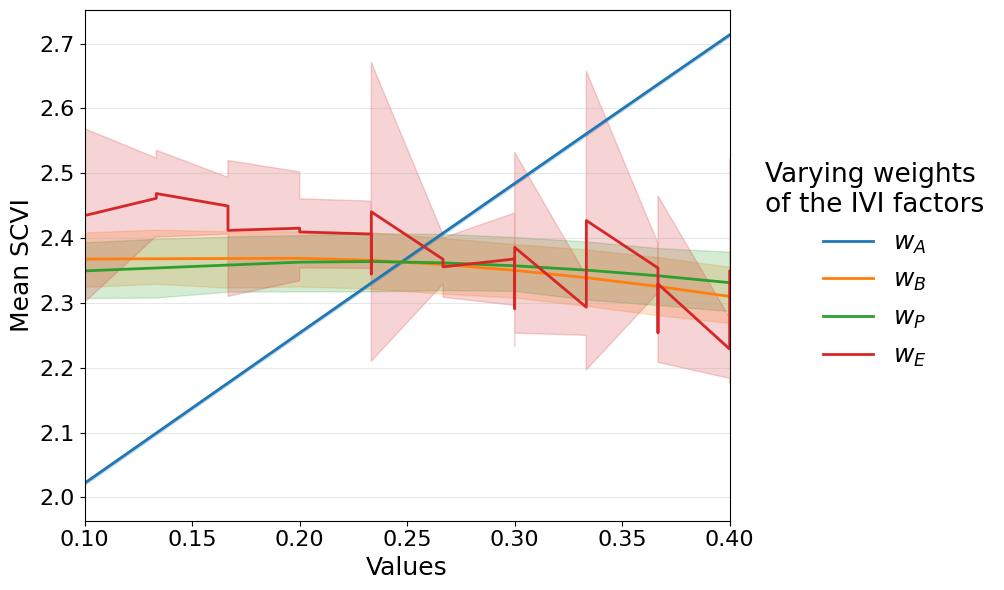}
        \label{fig:reddit_ivi_sensitivity}
    }
    \hfill
    \subfloat[Varying the weights of the ASI factors]{
        \includegraphics[width=0.45\textwidth]{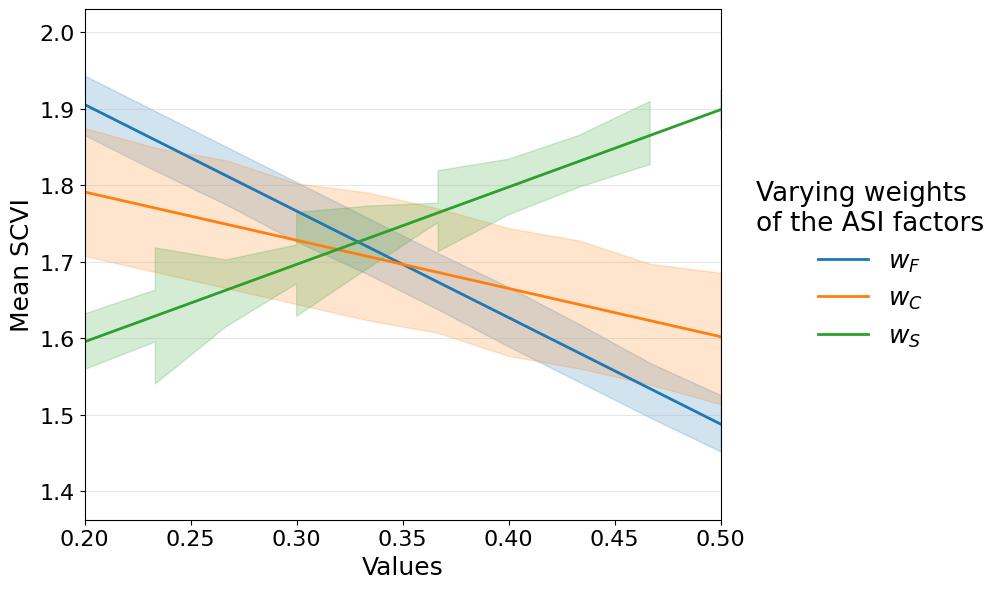}
        \label{fig:reddit_asi_sensitivity}
    }
    \caption{Sensitivity analysis of the Reddit dataset for IVI and ASI factors. Note that $w_A$, $w_B$, $w_P$, and $w_E$ are the weights for `Awareness,' `Behavioral,' `Psychological,' and `Experience' factors in the individual vulnerability index (IVI), respectively. $w_F$, $w_C$, and $w_S$ refer to `Frequency,' `Consequence,' and `Sophistication' in the Attack Security Index (ASI), respectively.  }
    \label{fig:reddit_sensitivity}
    \vspace{-5mm}
\end{figure*}

The sensitivity analysis on the iPoll dataset reveals key trends in the contributions of IVI and ASI components to SCVI variability, as illustrated in Fig.~\ref{fig:reddit_sensitivity}. Within the IVI, the {\it Awareness} factor (\(w_A\)) consistently increased SCVI scores, underscoring the exacerbating influence of limited user awareness. The {\it Experience} factor (\(w_E\)) exhibited a decreasing trend with noticeable fluctuations, indicating possible interactions with other components. The {\it Psychological} factor (\(w_P\)) showed a strong positive correlation with SCVI, while the {\it Behavioral} factor (\(w_B\)) exerted minimal positive influence, suggesting limited impact within the Reddit context.

For the ASI components in the iPoll dataset, the {\it Frequency} weight ($w_F$) was strongly negatively correlated with SCVI, implying that infrequent cyber-attacks contribute more to perceived vulnerability. Increasing the {\it Consequence} weight ($w_C$) consistently reduced SCVI, suggesting that preparedness for high-impact scenarios mitigates overall vulnerability. The {\it Sophistication} weight ($w_S$) showed a mild but variable positive association, indicating a context-dependent effect that may interact with other factors.

\begin{figure*}[t]
    \centering
    \subfloat[Varying the weights of the IVI factors]{
        \includegraphics[width=0.45\textwidth]{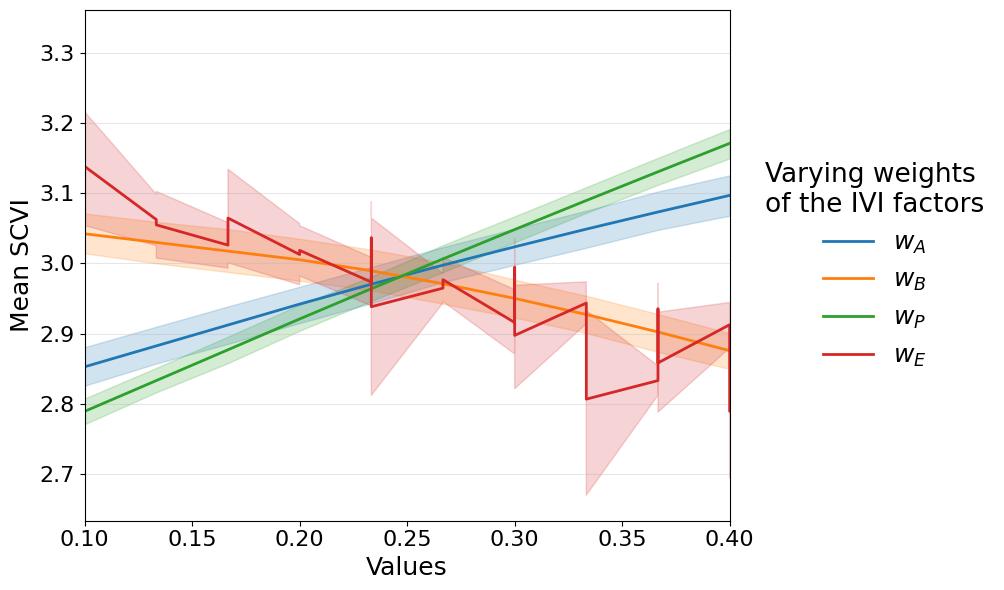}
        \label{fig:ipoll_ivi_sensitivity}
    }
    \hfill
    \subfloat[Varying the weights of the ASI factors]{
        \includegraphics[width=0.45\textwidth]{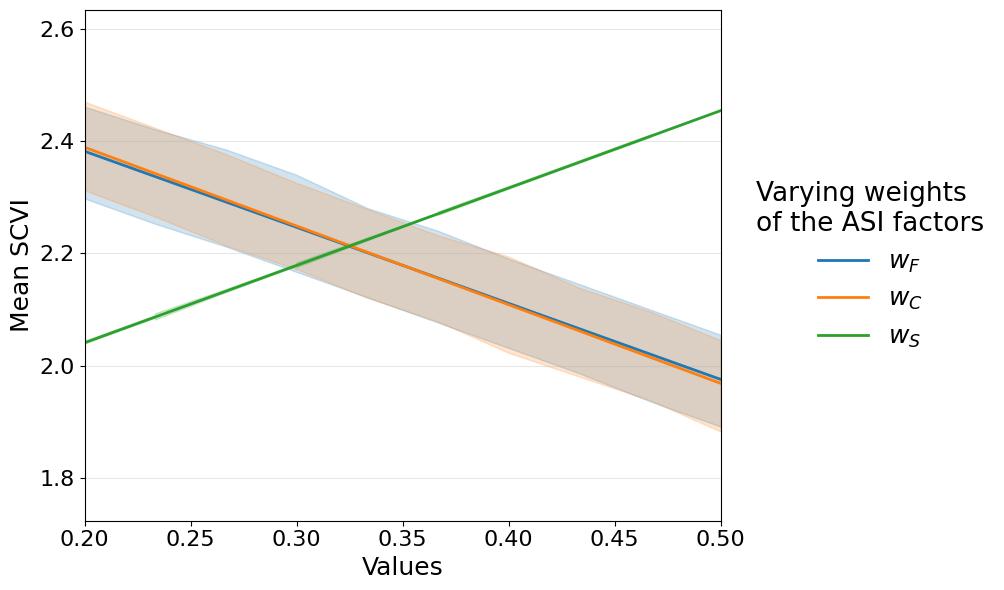}
        \label{fig:ipoll_asi_sensitivity}
    }
    \caption{Sensitivity analysis of the iPoll dataset for IVI and ASI factors. Note that $w_A$, $w_B$, $w_P$, and $w_E$ are the weights for `Awareness,' `Behavioral,' `Psychological,' and `Experience' factors in the individual vulnerability index (IVI), respectively. $w_F$, $w_C$, and $w_S$ refer to `Frequency,' `Consequence,' and `Sophistication' in the Attack Security Index (ASI), respectively.}
    \label{fig:ipoll_sensitivity}
\end{figure*}

The analysis of the Reddit dataset largely mirrored the iPoll findings, with some notable differences, as shown in Fig.~\ref{fig:ipoll_sensitivity}. The strong exacerbating effect of higher weights on {\it Awareness} (\(w_A\)) and the mitigating effect of {\it Experience} (\(w_E\)) on SCVI were consistent across both datasets. However, the {\it Psychological} factor (\(w_P\)) in the Reddit dataset exhibited a slight negative influence on SCVI, contrasting with the positive correlation observed in iPoll. Additionally, the {\it Behavioral} factor (\(w_B\)) showed a weak negative correlation with SCVI in Reddit, unlike the minimal positive influence noted in iPoll.

For ASI components, the {\it Frequency} factor (\(w_F\)) remained a significant driver of SCVI in both datasets, though its influence was stronger in Reddit. The {\it Consequence} factor (\(w_C\)) also consistently reduced SCVI but with a more pronounced effect in the Reddit dataset. Lastly, the {\it Sophistication} factor (\(w_S\)) had a clearer and more consistent positive impact in iPoll, whereas its influence in Reddit was less stable and more variable.

These analyses highlight the robust yet context-sensitive nature of SCVI components across datasets. {\it Awareness}, {\it Experience}, {\it Consequence}, and {\it Frequency} show consistent impacts across both iPoll and Reddit, while shifts in the magnitude and direction of {\it Psychological}, {\it Behavioral}, and {\it Sophistication} factors reflect dataset-specific characteristics. Overall, some SCVI components appear universal, whereas others are shaped by context. Future work will examine additional datasets and settings to strengthen the generalizability and robustness of SCVI as a comprehensive cyber vulnerability metric.

\begin{figure*}[t]
  \centering
\setcounter{subfigure}{0}
  \subfloat[Monte Carlo Analysis of SCVI Components in the iPoll Dataset: Primary and Secondary Peaks.]{    \includegraphics[width=0.45\textwidth, height=0.3\textwidth]{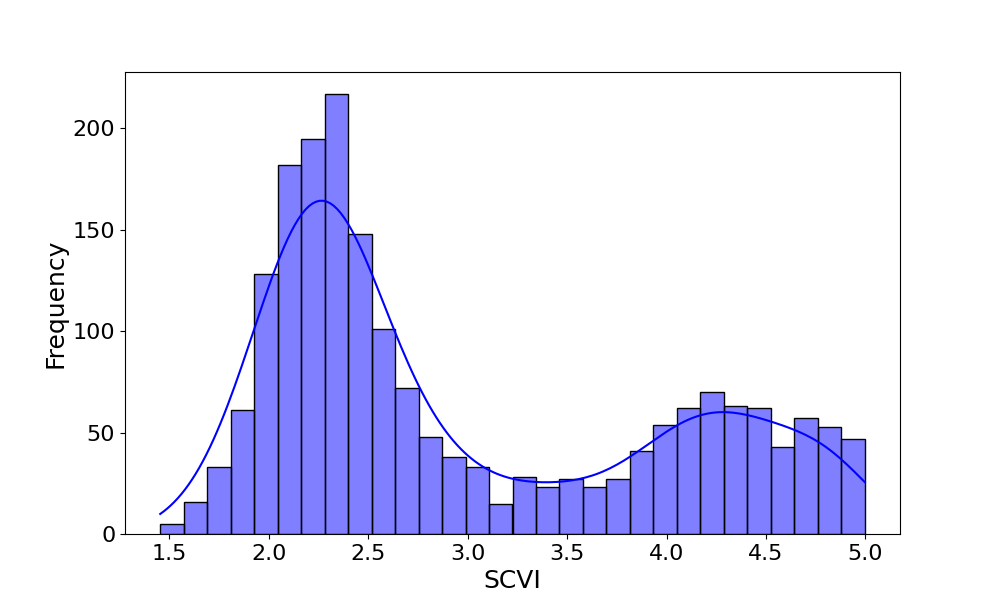} \label{fig:montecarlo_ipoll}}
  \hspace{2mm}
    \subfloat[Monte Carlo Analysis of SCVI Components in the Reddit Dataset: Low and High SCVI Outliers.]
    {\includegraphics[width=0.45\textwidth, height=0.3\textwidth]{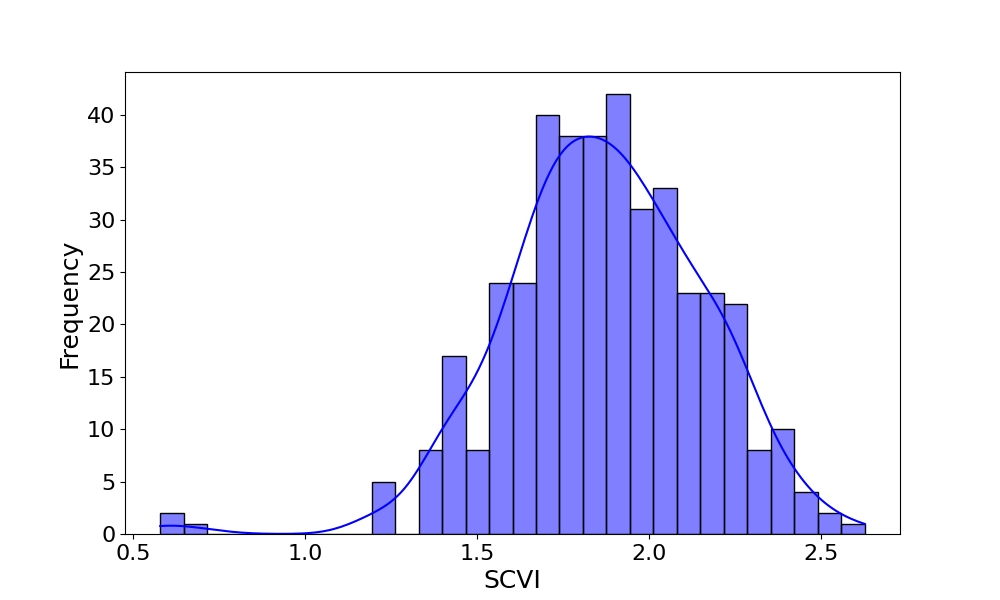}
    \label{fig:montecarlo_reddit}}
    \caption{Monte Carlo analysis of SCVI components in the iPoll and Reddit datasets}
\end{figure*}
\subsection{Effect of Weight Variability in SCVI}

SCVI values and weight configurations were aggregated across iterations to identify key patterns (Figs.~\ref{fig:montecarlo_ipoll} and~\ref{fig:montecarlo_reddit}).

\subsubsection{Analysis of iPoll Data}
Fig.~\ref{fig:montecarlo_ipoll} shows two dominant clusters, as follows.

\textbf{Primary Peak (Group 1)}: SCVI scores in this group were predominantly influenced by {\it Experience} (\(w_E\)) within the Individual Vulnerability Index (IVI) and {\it Sophistication} (\(w_S\)) within the Attack Severity Index (ASI). Experience had the highest contribution to IVI, with a mean of 0.421 and a standard deviation of 0.084. Similarly, Sophistication was the most significant ASI component, with a mean of 0.447 and a standard deviation of 0.095. These findings emphasize the combined importance of user experience and scam sophistication in shaping SCVI outcomes.

\textbf{Secondary Peak (Group 2)}: This group was primarily driven by {\it Awareness} (\(w_A\)) and {\it Psychological} factors (\(w_P\)) within IVI, alongside {\it Frequency} (\(w_F\)) in ASI. Awareness showed strong influence with a mean of 0.372 (Std: 0.026), while Frequency dominated ASI contributions (Mean: 0.436, Std: 0.033). In contrast to Group 1, Sophistication and Experience played minimal roles here, with SCVI more influenced by user awareness, psychological factors, and the rate of cyberattack encounters.

\subsubsection{Analysis of the Reddit Dataset}
Fig.~\ref{fig:montecarlo_reddit} shows two notable SCVI outlier groups.

\textbf{Low SCVI Outliers}: SCVI values ranged from 1.039 to 1.098, indicating low variability and high stability. {\it Behavioral} factors (\(w_B\)) were the primary contributors, ranging from 0.33 to 0.37, with {\it Experience} (\(w_E\)) also showing a meaningful contribution in certain cases. The weight distribution favored IVI, with \(\alpha\) values between 0.562 and 0.593, while ASI contributions were represented by \(\beta\) values ranging from 0.406 to 0.437. These patterns suggest that user behavior and prior experiences significantly shape vulnerability, and that SCVI remains a stable metric in these low-score contexts.

\textbf{High SCVI Outliers}: SCVI scores ranged from 2.182 to 2.211. {\it Experience} ($w_E$) and {\it Frequency} ($w_F$) were dominant drivers, both above 0.45, highlighting elevated risk for users with prior scam exposure and frequent threat encounters. Here, $\alpha$ ranged from 0.403 to 0.419, while $\beta$ ranged from 0.580 to 0.596, indicating stronger ASI influence. Overall, attack-specific features, especially frequency, elevate SCVI.

\subsubsection{Comparative Analysis of Both Datasets}
Both datasets underscore the dominant roles of {\it Experience} (\(w_E\)) and {\it Frequency} (\(w_F\)) in driving SCVI scores. However, notable differences emerge in how other factors influence vulnerability across contexts. In the Reddit dataset, {\it Behavioral} factors (\(w_B\)) played a key role in low SCVI cases, while {\it Frequency} had a more pronounced effect in high SCVI cases. This suggests that user behaviors are more critical in mitigating vulnerability, whereas the prevalence of cyberattacks becomes a stronger determinant of elevated SCVI scores.

In contrast, the iPoll dataset revealed stronger contributions from {\it Sophistication} (\(w_S\)) and {\it Awareness} (\(w_A\)), emphasizing the importance of scam complexity and user awareness in shaping vulnerability. This comparative analysis highlights the contextual nature of cyber risk, where individual behaviors and attack characteristics vary in their influence on SCVI depending on the data source. These differences reinforce the value of using diverse data inputs to ensure a comprehensive and adaptive assessment of cyber vulnerability.

\begin{figure}[h]
    \centering
    \includegraphics[width=0.3\textwidth]{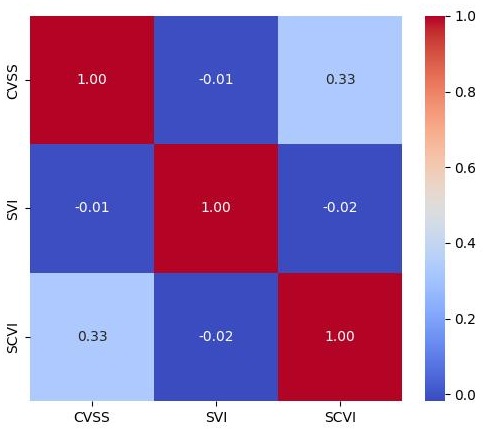}
    \caption{Spearman correlation heatmap across SCVI, SVI, and CVSS metrics.}
    \label{fig:spearman_correlation}
    \vspace{-5mm}
\end{figure}
Summarizing the insights above, while SCVI components exhibit universal trends, their relative contributions vary significantly across datasets, reflecting the contextual nature of cyber vulnerability. For example, the iPoll dataset highlights the dominant influence of {\it Sophistication} (\(w_S\)) and {\it Awareness} (\(w_A\)), emphasizing the role of attack complexity and user awareness in shaping risk. In contrast, the Reddit dataset reveals the critical impact of {\it Behavioral} factors (\(w_B\)) in low SCVI cases and {\it Frequency} (\(w_F\)) in high SCVI cases, illustrating how user behavior and the prevalence of attacks differentially influence vulnerability.

This variability highlights the importance of tailoring SCVI metrics to the specific characteristics of the dataset being analyzed. Future research should explore these dynamics further to enhance the adaptability and robustness of SCVI, enabling more precise and context-aware assessments of cyber vulnerability across diverse populations and platforms.

\subsection{SVI and CVSS Across Demographic Groups}
The SCVI metric shows strong potential for capturing individual-level vulnerabilities in social cyber contexts, outperforming traditional indices such as SVI and CVSS by integrating socio-demographic, behavioral, and cyber-specific factors. As shown in Fig.~\ref{fig:spearman_correlation}, correlation analysis indicates a moderate positive association between SCVI and CVSS (Spearman’s \(\rho = 0.33\), \(p = 0.0\)), suggesting alignment with CVSS in identifying technological vulnerabilities.

In contrast, SCVI exhibits a near-zero correlation with SVI (\(\rho = -0.01\), \(p = 0.4836\)), highlighting the divergence between SCVI’s cyber-focused vulnerability assessment and SVI’s emphasis on broader socio-environmental conditions. This contrast underscores SCVI’s distinct value in addressing gaps left by conventional metrics and its relevance in cyber-specific risk assessment across diverse demographic groups.

\begin{figure*}[t]
    \centering
    \subfloat[CVSS score distribution]{
        \includegraphics[width=0.3\textwidth, height=0.25\textwidth]{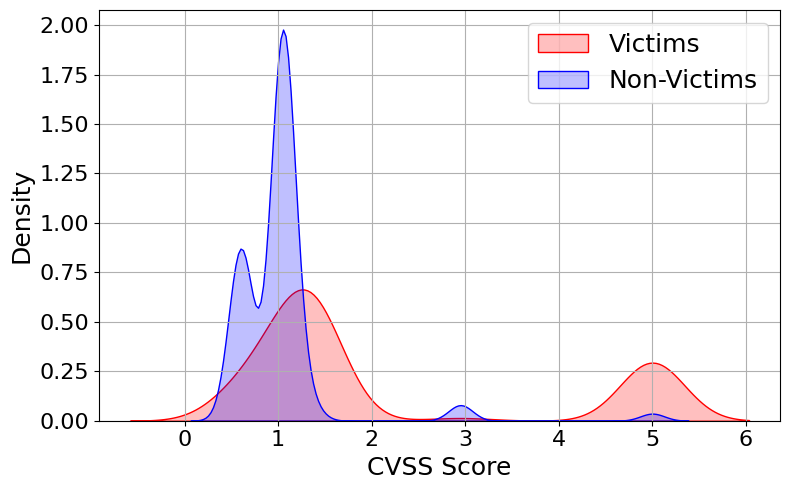}
        \label{fig:cvss}
    }
    \hspace{1mm}
    \subfloat[SVI score distributions]{
        \includegraphics[width=0.3\textwidth, height=0.25\textwidth]{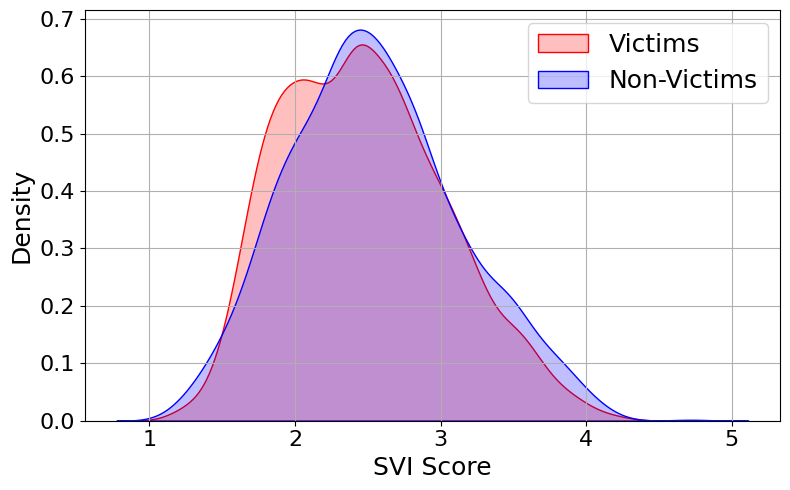}
        \label{fig:svi}
    }
    \hspace{1mm}
    \subfloat[SCVI score distribution]{
    \includegraphics[width=0.3\textwidth, height=0.25\textwidth]{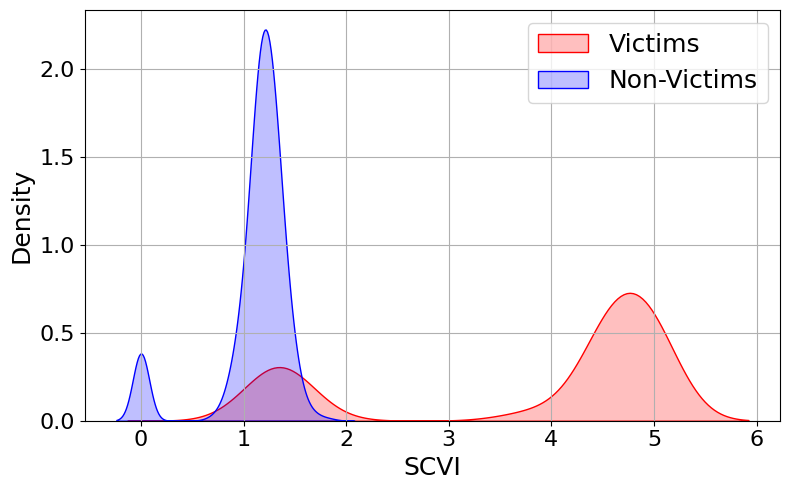}
        \label{fig:scvi}
    }
    \caption{Comparison of CVSS, SVI, and SCVI score distributions by victim class using the Ipoll dataset. SCVI shows greater discriminative power between victim and non-victim groups than CVSS and SVI.}
    \label{fig:kde}
    \vspace{-5mm}
\end{figure*}
Based on the density plots in Fig.~\ref{fig:kde}, which compare CVSS, SVI, and SCVI across victim and non-victim groups, a clear narrative emerges regarding the strengths and limitations of each metric in distinguishing cyber vulnerability.

Fig.~\ref{fig:cvss} presents the distribution of CVSS (Common Vulnerability Scoring System) scores. Victim scores exhibit a bimodal pattern with peaks near 1.5 and 5, whereas non-victims are more tightly clustered around lower scores, particularly near 1. This indicates that while CVSS captures some aspects of technical risk, it has limited discriminative power in separating victims from non-victims. The overlapping regions in the distributions reflect CVSS’s inability to incorporate human or contextual factors critical to real-world exploitation.

As shown in Fig.~\ref{fig:svi}, SVI scores provide limited separation between groups. Victim and non-victim populations show unimodal distributions centered at 2--3 with overlap, suggesting that while SVI captures broader social risks (e.g., poverty, education, housing), it lacks the granularity to predict individual-level cyber victimization.

In contrast, SCVI distributions shown in Fig.~\ref{fig:scvi} reveal a pronounced separation. Non-victims cluster tightly around low SCVI scores near 1, while victims exhibit a bimodal distribution with a prominent peak around 4.8. This distinct contrast highlights SCVI’s superior ability to distinguish between victim and non-victim groups. By integrating socio-demographic, behavioral, and cyber-specific risk factors, such as exposure and attack frequency, SCVI captures the intersectional nature of cyber vulnerability. It demonstrates that individuals most at risk are not merely socially or technically vulnerable, but rather positioned at the confluence of both dimensions.

Analysis across gender, race-ethnicity, and age groups reveals consistent patterns in vulnerability metrics. As shown in Table~\ref{tab:svi-cvss-scvi} (GENDER Table), the male group records the highest average values across all three indices: SVI (2.57), CVSS (3.52), and SCVI (3.47). While the SVI difference between male and female groups is modest (2.36\%), the disparities in CVSS and SCVI are more pronounced, 9.84\% and 16.88\%, respectively, possibly reflecting greater exposure to higher-risk technological environments among males.

Table~\ref{tab:svi-cvss-scvi} (RACE-ETHNICITY Table) reports vulnerability scores across racial and ethnic groups. The White, non-Hispanic group shows the lowest values (SVI 2.25; CVSS 3.02), while the Hispanic group exhibits the highest scores: SVI (3.72), CVSS (3.79), and SCVI (4.06). SCVI consistently exceeds CVSS and SVI, indicating a need for tailored cybersecurity interventions for Hispanic populations. Lower scores for White, non-Hispanic groups may reflect greater access to protective resources, education, or institutional support.

Overall, Table~\ref{tab:svi-cvss-scvi} reveals a clear trend: SVI scores are generally lower than both CVSS and SCVI scores. The only exceptions are the Black, non-Hispanic group, where SVI slightly exceeds CVSS by 0.3\%, and the Other, non-Hispanic group, where SVI exceeds CVSS by 10\%. No group showed SVI as the highest among the three indices, underscoring its limits in capturing cyber-specific dimensions of vulnerability.

\begin{table*}[t]
\centering
\small
\caption{\sc Distribution of SVI, CVSS, and SCVI Metrics by Demographic Attributes}
\label{tab:svi-cvss-scvi}

\subfloat[\sc Gender\label{tab:gender}]{
\begin{minipage}[t]{0.3\textwidth}\centering
\begin{tabular}{|l|c|c|c|}
\hline
Gender & SVI & CVSS & SCVI \\ \hline
Female & 2.51 & 3.19 & 2.93 \\ \hline
Male   & 2.57 & 3.52 & 3.47 \\ \hline
\end{tabular}
\end{minipage}
}
\subfloat[\sc Race-Ethnicity\label{tab:race_ethnicity}]{
\begin{minipage}[t]{0.3\textwidth}\centering
\begin{tabular}{|l|c|c|c|}
\hline
Race-Ethnicity & SVI & CVSS & SCVI \\ \hline
White, non-Hispanic & 2.25 & 3.31 & 3.02 \\ \hline
Black, non-Hispanic & 3.34 & 3.34 & 3.40 \\ \hline
Other, non-Hispanic & 3.01 & 3.74 & 3.33 \\ \hline
2+, non-Hispanic    & 2.89 & 3.32 & 3.14 \\ \hline
Asian, non-Hispanic & 2.90 & 3.32 & 3.02 \\ \hline
Hispanic            & 3.72 & 4.06 & 3.79 \\ \hline
\end{tabular}
\end{minipage}
}
\hfill
\subfloat[\sc Age Groups\label{tab:age_groups}]{
\begin{minipage}[t]{0.3\textwidth}\centering
\begin{tabular}{|l|c|c|c|}
\hline
Age Group & SVI & CVSS & SCVI \\ \hline
18--24 & 3.85 & 4.69 & 4.14 \\ \hline
25--29 & 3.11 & 3.59 & 3.50 \\ \hline
30--44 & 3.31 & 3.33 & 3.11 \\ \hline
45--49 & 2.74 & 3.82 & 3.45 \\ \hline
50--54 & 2.42 & 3.81 & 3.45 \\ \hline
55--64 & 2.25 & 3.24 & 2.95 \\ \hline
65+    & 2.12 & 2.82 & 2.63 \\ \hline
\end{tabular}
\end{minipage}
}

\end{table*}

In the age group analysis presented in Table~\ref{tab:svi-cvss-scvi} (AGE-GROUP Table), younger demographics (ages 18--24 and 25--29) exhibit the highest overall vulnerability, with SCVI scores consistently exceeding those of CVSS and SVI. This pattern suggests greater susceptibility to social cyber vulnerabilities, likely driven by higher levels of technology use and online engagement. As age increases, all three indices, SCVI, CVSS, and SVI, generally decrease, with CVSS scores leading among middle-aged and older groups. This may reflect a shift from behavioral risks toward more technical exposure.

Interestingly, a slight uptick in all metrics is observed for the 45--49 and 50--54 age groups, potentially indicating increased digital engagement but lower scam awareness compared to adjacent age cohorts. Across all age groups, SVI consistently registers as the lowest metric, further underscoring its limited sensitivity to cyber-specific factors.

These findings validate SCVI as a robust, comprehensive metric that outperforms traditional indices in capturing nuanced social cyber vulnerability. SCVI consistently yields the highest scores across demographic categories, highlighting its ability to identify risks linked to technological behaviors, demographic factors, and cyber threat exposure. This distinction reinforces SCVI’s value for assessing and addressing individual-level vulnerabilities, particularly for groups with elevated susceptibility to cyber threats.

\subsection{Regional Disparities in Cyber Vulnerability: Analysis of SCVI Across U.S. States Using the iPoll Dataset}

\begin{figure}[t]
    \centering
    \includegraphics[width=0.45\textwidth]{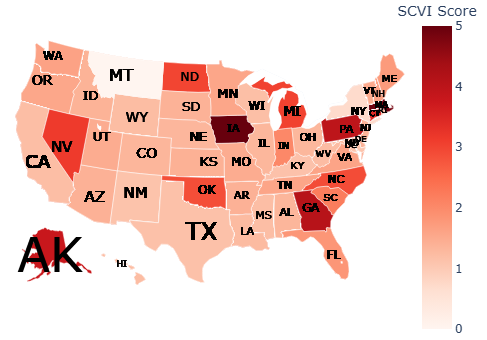}
    \caption{Heatmap of SCVI across the United States where color intensity reflects sample size to indicate data reliability.}
    \label{fig:heatmap}
    \vspace{-5mm}
\end{figure}
The SCVI computed from the iPoll dataset enables analysis of regional disparities in cyber vulnerability across the United States. The dataset reports mean SCVI scores, confidence intervals, and sample sizes by state, providing insight into the geographic distribution of cyber threats and the potential effectiveness of local cybersecurity measures.

Data visualization used a heatmap (Fig.~\ref{fig:heatmap}), with color intensity scaled by sample size to reflect data reliability. Table~III (see the supplement document) summarizes state-level SCVI scores, sample sizes, and confidence intervals.

States such as Alaska, Rhode Island, and Nevada exhibit higher mean SCVI scores alongside smaller sample sizes, indicating a potential elevation in cyber vulnerability that may be shaped by regional factors. The wide confidence intervals for these states underscore significant uncertainty in SCVI estimates, primarily due to limited data availability.

Conversely, populous states such as California, Texas, and New York show lower and more stable SCVI scores, suggesting reduced cyber vulnerability or more effective cybersecurity practices; their larger samples also support more reliable risk estimates. Discrepancies between states with similar sample sizes (e.g., Connecticut vs.\ Pennsylvania) indicate that socio-economic or infrastructural factors contribute to cyber vulnerability. These findings align with regional patterns reported by the FTC 2020 Data Book~\cite{FTC2021}.

An inverse relationship between mean SCVI scores and sample sizes was observed in several states, indicating that smaller samples may capture localized extremes or anomalies not reflected in broader datasets. This reinforces the importance of cautious interpretation and the need for additional investigation in regions with sparse data.

Overall, this analysis underscores the need for region-specific cyber vulnerability assessments and tailored cybersecurity policies that address state-level challenges. Improved data collection and larger sample sizes will strengthen SCVI reliability and inform more effective regional and national cybersecurity strategies.

\section{Discussions: Key Findings and Limitations}

\subsection{Key Findings}
Findings from the iPoll and Reddit datasets reveal several consistent trends in user vulnerabilities and attack characteristics. \textbf{\em First}, many individuals report low self-assessed vulnerability (IVI) yet face high-severity scams (ASI), indicating a mismatch between perceived and actual risk and motivating targeted awareness and coping interventions for high-impact scams (e.g., phishing, romance, and investment fraud). \textbf{\em Second}, among IVI components, {\it psychological factors} (e.g., impulsivity, credulity, emotional distress) and {\it experience} (e.g., prior victimization) are particularly influential, suggesting that effective mitigation should combine technical guidance with resilience- and emotion-aware education. \textbf{\em Third}, SCVI’s dual-layer integration of IVI and ASI provides finer-grained risk assessment than CVSS or SVI: SCVI aligns moderately with CVSS (\(r=0.33\)) yet is largely independent of SVI (\(r\approx -0.01\)), enabling intervention design tailored to scam types and the behavioral/psychological mechanisms they exploit. \textbf{\em Fourth}, sensitivity and weight-variability analyses show contextual adaptability: {\it Awareness} (\(w_A\)) consistently elevates SCVI, while {\it Experience} (\(w_E\)) is interaction-dependent; {\it Psychological} effects (\(w_P\)) are stronger in iPoll but slightly negative in Reddit, whereas {\it Frequency} (\(w_F\)) is more influential in Reddit and {\it Sophistication} (\(w_S\)) is more consistent in iPoll. \textbf{\em Fifth}, geographic analysis (iPoll) indicates state-level disparities driven by local risk environments and data availability: small-sample states (e.g., Alaska, Rhode Island) show higher SCVI and wider confidence intervals, while populous states (e.g., California, Texas) exhibit lower, more stable SCVI, supporting region-specific cyber policy and public awareness campaigns.

\subsection{Limitations}

Despite the robustness and novelty of the SCVI framework, several limitations merit consideration. \textbf{\em First}, both the iPoll survey~\cite{ipoll-dataset2020} and Reddit scam reports~\cite{ftc_2020} rely on self-reported data, which can introduce recall bias and under-reporting, reducing reliability. \textbf{\em Second}, sampling and self-selection effects may skew coverage: iPoll may underrepresent groups with limited internet access or language barriers, while Reddit is biased toward more digitally literate users. \textbf{\em Third}, state-level sample size variability can yield wide confidence intervals, limiting the reliability of regional comparisons and motivating broader, more balanced data collection. \textbf{\em Fourth}, uniform component weighting (e.g., equal IVI/ASI emphasis) may oversimplify complex interactions; data-driven or adaptive weighting could better capture nuanced vulnerability dynamics. \textbf{\em Fifth}, cross-sectional datasets limit causal inference and trend analysis, and sociocultural differences in persuasion and trust may constrain generalizability, motivating longitudinal and cross-cultural validation. \textbf{\em Sixth}, the evolving threat landscape requires periodic updates to SCVI to reflect changes in scam tactics and social engineering methods. Collectively, these limitations motivate refinements that improve SCVI’s adaptability, scalability, and robustness as a comprehensive metric for social cyber vulnerability assessment.

\section{Conclusions \& Future Work}

In this work, we introduced the Social Cyber Vulnerability Index (SCVI), an interpretable and uncertainty-aware socio-cyber risk metric that bridges user susceptibility and attack severity by fusing an Individual Vulnerability Index (IVI) (awareness/knowledge, behavior, psychological factors, and experience) with an Attack Severity Index (ASI) (frequency, consequence, and sophistication). We operationalized SCVI across heterogeneous modalities, a nationally scoped iPoll survey (4,596 U.S.\ adults) and 450 Reddit \texttt{r/scams} reports (2016--2024), showing that SCVI can be computed from both structured questionnaires and CI-driven feature extraction from social-media narratives. We quantified robustness via sensitivity analysis and 10,000-iteration Monte Carlo simulations, demonstrating stable behavior under plausible weight variability while revealing context-dependent drivers. Comparisons with CVSS and SVI highlighted SCVI’s distinct socio-technical coverage (moderate alignment with CVSS; near-independence from SVI) and stronger discriminative power for separating victim and non-victim groups, enabling more reliable identification of high-risk populations. Finally, SCVI surfaced demographic and regional disparities and informed prioritization of targeted interventions across scam typologies, providing a practical foundation for scalable socio-cyber vulnerability assessment under evolving, AI-enabled threats.

To address current limitations, \textbf{future work} will focus on improving intervention relevance, measurement rigor, and generalizability. SCVI can enable targeted interventions by jointly modeling user susceptibility and high-impact threats, especially when users underestimate risk relative to encountered scams. Improving data reliability is essential; behavioral telemetry, passive security signals, or third-party validation can reduce self-report bias and strengthen robustness. Broader demographic and geographic coverage, including underrepresented groups (e.g., older adults and non-English speakers), will improve inclusivity, while increasing samples in sparse regions and using stratified designs or region-specific weighting can reduce regional variance. Longitudinal studies can capture how awareness, behavior, and attacker tactics evolve over time, enabling proactive risk management. Finally, SCVI should use data-driven weighting and regular updates to address emerging typologies (e.g., AI-enabled phishing and deepfakes) and incorporate cultural differences in trust cues and persuasion to improve global relevance.

\bibliographystyle{elsarticle-harv} 
\bibliography{ref}

\end{document}


\title{Supplement Materials: Assessing Socio-Cyber Vulnerability Using Survey and Social Media Data}

\author{Shutonu~Mitra,
        Tomas~Neguyen,
        Qi~Zhang,
        Hossein~Salemi,
        Fengxiu~Zhang,
        Michin~Hong,
        Chang-Tien~Lu,~\IEEEmembership{Fellow,~IEEE,}
        Hemant~Purohit,
        Jin-Hee~Cho,~\IEEEmembership{Senior~Member,~IEEE,}%
\thanks{Shutonu Mitra (email: \href{mailto:mshutonu@vt.edu}{mshutonu@vt.edu}), Tomas Neguyen (email: \href{mailto:thomasn03@vt.edu}{thomasn03@vt.edu}), Qi Zhang (email: \href{mailto:qiz21@vt.edu}{qiz21@vt.edu}), Chang-Tien Lu (email: \href{mailto:clu@vt.edu}{clu@vt.edu}), and Jin-Hee Cho (email: \href{mailto:jicho@vt.edu}{jicho@vt.edu}) are with Virginia Tech, Blacksburg, VA, USA.}%
\thanks{Hossein Salemi (email: \href{mailto:hsalemi@gmu.edu}{hsalemi@gmu.edu}), Fengxiu Zhang (email: \href{mailto:fzhang22@gmu.edu}{fzhang22@gmu.edu}), and Hemant Purohit (email: \href{mailto:hpurohit@gmu.edu}{hpurohit@gmu.edu}) are with George Mason University, Fairfax, VA, USA.}%
\thanks{Michin Hong (email: \href{mailto:hongmi@iu.edu}{hongmi@iu.edu}) is with Indiana University, Bloomington, IN, USA.}%
}

\markboth{IEEE Transactions on Emerging Topics in Computational Intelligence,~Vol.~xx, No.~x, x-month~2026}{}


\maketitle

\appendix

\noindent \textbf{IVI feature extraction from iPoll.}
Table~\ref{tab:ivi-calculation} summarizes how we operationalize the IVI components from iPoll survey items. For each IVI subcomponent (gray headers), we map categorical responses to ordinal scores so that larger values consistently indicate \emph{higher} individual vulnerability (e.g., lower familiarity/knowledge, riskier behaviors, higher trust or impulsivity). ``DON'T KNOW,'' ``SKIPPED,'' and ``REFUSED'' responses are treated as missing and excluded from aggregation. For each respondent, item-level scores are combined within each subcomponent (e.g., $A_{1,k}$, $P_{2,k}$) to obtain the corresponding factor values used in Eqs.~(2) and~(3).

\vspace{1mm}
\noindent \textbf{ASI feature extraction from iPoll.}
Table~\ref{tab:asi_calculation-1} details the response-to-score mappings used to derive ASI-related quantities from iPoll. We use self-reported exposure and victimization questions as proxies for attack \emph{frequency}, reported losses and distress to quantify \emph{consequences}, and scam-related realism indicators to approximate \emph{sophistication}. As in Table~\ref{tab:ivi-calculation}, responses are mapped to ordinal scores with higher values indicating more severe attack conditions, while ``DON'T KNOW,'' ``SKIPPED,'' and ``REFUSED'' are ignored in aggregation. The resulting factor values ($F_{i,k}$, $C_{i,k}$, $S_{i,k}$) are then used to compute $\mathrm{ASI}_{i,k}$ via Eqs.~(4)--(5).

\vspace{1mm}
\noindent \textbf{State-level summary statistics.}
Table~\ref{tab:state_summary} reports state-wise aggregates of IVI, ASI, and SCVI. For each state, we list the sample size and the mean scores computed over respondents in that state, along with the corresponding confidence interval (CI) bounds for the mean SCVI. Reporting sample size alongside CIs is important because states with small samples naturally yield wider intervals; therefore, state comparisons should be interpreted with appropriate caution when $n$ is limited.

\begin{table*}[h!]
\centering
\caption{\sc Feature Extraction for Individual Vulnerability Index (IVI) from the iPoll Dataset}
\label{tab:ivi-calculation}
\footnotesize
\begin{tabular}{|P{6cm}|P{6cm}|}
\hline
 {\bf Related Questions}  & {\bf Response and Score Mapping}\\
\hline
\hline

\rowcolor{gray!15}\multicolumn{2}{|c|}{\bf Lack of Awareness $A_{1,k}$}     \\

\hline
Q7. Generally, how concerned, if at all, are you that you and/or a family member may fall victim to a scam? & Very concerned: 0, Somewhat concerned: 1, Not too concerned: 2, Not at all concerned: 3, DON’T KNOW/ SKIPPED ON WEB/REFUSED: ignore \\
\hline
Q8. How familiar were you with online romance scams; Q14: Grandparent scams? Q21: Government impostor scams; Q28 Census scams & Very: 0, Somewhat: 1, A little: 2, Not at all: 3, DON’T KNOW/ SKIPPED ON WEB/REFUSED: ignore \\
\hline
\rowcolor{gray!15}\multicolumn{2}{|c|}{\bf Lack Knowledge of Protect Measure $A_{2,k}$}     \\

\hline
Q37. Is Caller ID a reliable way to know where a call comes from? & True: 0; False: 5, Not sure: 3, SKIPPED ON WEB/REFUSED: ignore \\
\hline
Q38. When surfing the internet, it is always safe to interact with a website as long as the website has a locked box icon that indicates it is HTTPS secured. & Same as above \\
\hline
Q39. The IRS can call you about back taxes you may owe without sending you a written notice first. & True: 5; False: 0, Not sure: 2, SKIPPED ON WEB/REFUSED: ignore \\
\hline
Q40. The Social Security Administration will contact you directly, either by phone or email, if there is a problem with your Social Security benefits. & Same as above \\
\hline
\rowcolor{gray!15}\multicolumn{2}{|c|}{\bf Frequency of behaviors increasing the risk of attack $B_{1,k}$}     \\
\hline
Q1. Not including time you spend participating in online surveys, how often do you typically go online or access the Internet, including sending or receiving email? & Daily: 5, Several times a week: 4, several times a month: 3, Once a month: 2, Less than once a month: 1, Never: 0 \\
\hline
Q2. How often, if at all, do you use the Internet to do the following activities & Daily: 5, Several times a week: 4, several times a month: 3, Once a month: 2, Less than once a month: 1, Never: 0 \\
\hline
Q3. Have you ever done any of the following to meet potential dates or romantic partners at any point or time in your life? & Yes: 5, No: 0, Not sure: 3 \\
\hline

\rowcolor{gray!15}\multicolumn{2}{|c|}{\bf Trust level related to attack $P_{1,k}$}     \\
\hline
Q6. How well do the following statements describe you? (Overall, I expect more good things to happen to me than bad; I sympathize with others' feelings; I am a trusting person; Overall, I am pleased with my life.) & very well: 5, Somewhat: 3, Not at all: 0 \\
\hline
Q6. How well do the following statements describe you? (I find it difficult to get emotionally close to others; I worry a lot.) & very well: 0, Somewhat: 3, Not at all: 5 \\
\hline
\rowcolor{gray!15}\multicolumn{2}{|c|}{\bf Risk perception and impulsivity $P_{2,k}$}     \\
\hline
Q5. Have you ever developed a romantic relationship with someone whom you have never met in person? & Yes: 5, No: 0, DON’T KNOW/ SKIPPED ON WEB/REFUSED: ignore \\
\hline
Q6. How well do the following statements describe you? (I tend to get involved in things that I later wish I could get out of; I tend to make up my mind quickly; I feel uneasy in social settings & Very well: 5, Somewhat: 3, Not at all: 0 \\
\hline
\end{tabular}
\end{table*}

\begin{table*}[ht]
\centering
\caption{Feature Extraction for the Attack Severity Index (ASI) from the iPoll Dataset.}
\label{tab:asi_calculation-1}
\footnotesize
\begin{tabular}{|p{6cm}|p{6cm}|}
\hline
\rowcolor{gray!15}\multicolumn{2}{|c|}{\bf Past Experiences $E_{1,k}$}     \\
\hline
Q4. Thinking of the dates or romantic partners that you have met first online, have any of them ever done the following? (Lied about themselves, ask for money, etc.) & Yes: 5, No: 0, Not sure: 3 \\
\hline
Q9. To the best of your knowledge, have you ever been a target of a romance scam, Q15, or a Grandparent scam, Q22? Government impostor scams; Q29. Census scams; Q35. Identity theft&  Yes: 5, No: 0, Not sure: 3, SKIPPED ON WEB/REFUSED: ignore \\
\hline
Q12. To the best of your knowledge, has anyone you know ever been a target of a romance scam? Q19: Grandparent scam. Q26. government impostor scams. Q33. Census scams & Yes: 0, No: 5, Not sure: 3, SKIPPED ON WEB/REFUSED: ignore \\
\hline
Q13. Did the person lose any money or suffer other financial losses due to the romance scam?  Q20: Grandparent scam. Q27. government impostor scams. Q34. Census scams  & Same as above \\
\hline
Q36. Approximately when did you experience identity theft?  & less than a year: 1, 1-2 year: 2, 3-5 year: 3, 5-9 year: 4, more than 10 years: 5, don't know/skipped/refused: ignore \\
\hline
\rowcolor{gray!15}\multicolumn{2}{|c|}{\bf Responses to past incidents $E_{2,k}$}     \\
\hline
Q10. Have you ever lost money or suffered other financial losses due to a romance scam? Q17: Grandparent scam. Q24. government impostor scams; Q31. Census scams  & Yes: 5, No: 0, DON’T KNOW/ SKIPPED ON WEB/REFUSED: ignore \\
\hline
Q11. Have you ever experienced health problems or emotional distress due to a romance scam? Q18: Grandparent scam. Q25. government impostor scams; Q32. Census scams  & Yes, health problems only: 3, Yes, emotional distress only: 3, both: 5, no: 0, DON’T KNOW/ SKIPPED ON WEB/REFUSED: ignore \\
\hline

\hline
\rowcolor{gray!15}\multicolumn{2}{|c|}{\bf Frequency Factor} \\
\hline
Q9. ``To the best of your knowledge, have you ever been a target of a romance scam?''    
& Yes: 5, No: 0, Not sure: 1, SKIPPED/REFUSED: ignore \\
\hline
Q12. ``To the best of your knowledge, has anyone you know ever been a target of a romance scam?''   
& Yes: 5, No: 0, Not sure: 1, SKIPPED/REFUSED: ignore \\
\hline
Q15. ``To the best of your knowledge, have you ever been a target of a grandparent scam?''  
& Yes: 5, No: 0, Not sure: 1, SKIPPED/REFUSED: ignore \\
\hline
Q19. ``To the best of your knowledge, has anyone you know ever been a target of a grandparent scam?''  
& Yes: 5, No: 0, Not sure: 1, SKIPPED/REFUSED: ignore \\
\hline
Q22. ``To the best of your knowledge, have you ever been a target of a government impostor scam?''   
& Yes: 5, No: 0, Not sure: 1, SKIPPED/REFUSED: ignore \\
\hline
Q26. ``To the best of your knowledge, has anyone you know ever been a target of a government impostor scam?''   
& Yes: 5, No: 0, Not sure: 1, SKIPPED/REFUSED: ignore \\
\hline
Q29. ``To the best of your knowledge, have you ever been a target of a Census scam?'' 
& Yes: 5, No: 0, Not sure: 1, SKIPPED/REFUSED: ignore \\
\hline
Q33. ``To the best of your knowledge, has anyone you know ever been a target of a Census scam?''  
& Yes: 5, No: 0, Not sure: 1, SKIPPED/REFUSED: ignore \\
\hline

\end{tabular}
\end{table*}

\begin{table*}[ht]
\centering
\caption*{\sc Table II (cont'd): Feature Extraction for the Attack Severity Index (ASI) from the iPoll Dataset.}
\label{tab:asi_calculation}
\footnotesize
\begin{tabular}{|p{6cm}|p{6cm}|}
\hline
\rowcolor{gray!15}\multicolumn{2}{|c|}{\bf Consequence Factor} \\
\hline
Q10. ``Have you ever lost money or suffered other financial losses due to a romance scam?''    
& Yes: 5, No: 0, Not sure: 1, SKIPPED/REFUSED: ignore \\
\hline
Q11. ``Have you ever experienced any health problems or emotional distress due to a romance scam?''    
& Yes, health only: 4; Yes, emotional distress only: 4; Yes, both: 5; No: 0, SKIPPED/REFUSED: ignore \\
\hline
Q13. ``Did the person lose any money or suffer other financial losses due to the romance scam?''    
& Yes: 5, No: 0, Not sure: 1, SKIPPED/REFUSED: ignore \\
\hline
Q17. ``Have you ever lost money or suffered other financial losses due to a grandparent scam?''    
& Yes: 5, No: 0, Not sure: 1, SKIPPED/REFUSED: ignore \\
\hline
Q18. ``Have you ever experienced any health problems or emotional distress due to a grandparent scam?''    
& Yes, health only: 4; Yes, emotional distress only: 4; Yes, both: 5; No: 0, SKIPPED/REFUSED: ignore \\
\hline
Q20. ``Did the person lose any money or suffer other financial losses due to the grandparent scam?''    
& Yes: 5, No: 0, Not sure: 1, SKIPPED/REFUSED: ignore \\
\hline
Q24. ``Have you ever lost money or suffered other financial losses due to a government impostor scam?''    
& Yes: 5, No: 0, Not sure: 1, SKIPPED/REFUSED: ignore \\
\hline
Q25. ``Have you ever experienced any health problems or emotional distress due to a government impostor scam?''    
& Yes, health only: 4; Yes, emotional distress only: 4; Yes, both: 5; No: 0, SKIPPED/REFUSED: ignore \\
\hline
Q27. ``Did the person lose any money or suffer other financial losses due to the government impostor scam?''    
& Yes: 5, No: 0, Not sure: 1, SKIPPED/REFUSED: ignore \\
\hline
Q31. ``Have you ever lost money or suffered other financial losses due to a Census scam?''    
& Yes: 5, No: 0, Not sure: 1, SKIPPED/REFUSED: ignore \\
\hline
Q32. ``Have you ever experienced any health problems or emotional distress due to a Census scam?''    
& Yes, health only: 4; Yes, emotional distress only: 4; Yes, both: 5; No: 0, SKIPPED/REFUSED: ignore \\
\hline
Q34. ``Did the person lose any money or suffer other financial losses due to the Census scam?''    
& Yes: 5, No: 0, Not sure: 1, SKIPPED/REFUSED: ignore \\
\hline

\rowcolor{gray!15}\multicolumn{2}{|c|}{\bf Sophistication Factor} \\
\hline
Q10, Q13, Q17, Q20, Q24, Q27, Q31, Q34 
& Yes: 5, No: 0, Not sure: 1, SKIPPED/REFUSED: ignore \\
\hline
\end{tabular}
\end{table*}


\begin{table*}[t]
\centering
\caption{\sc State-wise Summary of Mean IVI, ASI, SCVI, and Confidence Intervals  (CIs)}
\label{tab:state_summary}
\footnotesize
\begin{tabular}{@{}lcccccc@{}}
\toprule
\textbf{State} & \textbf{Sample Size} & \textbf{Mean IVI} & \textbf{Mean ASI} & \textbf{Mean SCVI} & \textbf{CI Lower} & \textbf{CI Upper} \\ \midrule
Alabama         & 21  & 2.2017 & 1.1519 & 1.7487 & 1.2129 & 2.2844 \\
Alaska          & 2   & 2.6621 & 2.5000 & 2.6511 & 0.4127 & 4.8894 \\
Arizona         & 52  & 2.4559 & 1.3683 & 1.9859 & 1.6792 & 2.2927 \\
Arkansas        & 14  & 2.2154 & 1.5321 & 1.8738 & 1.4236 & 2.3240 \\
California      & 197 & 2.3034 & 1.4641 & 1.9769 & 1.8126 & 2.1411 \\
Colorado        & 43  & 2.3126 & 1.2260 & 1.9192 & 1.5396 & 2.2989 \\
Connecticut     & 648 & 2.2852 & 1.0578 & 1.7606 & 1.6762 & 1.8451 \\
Delaware        & 8   & 2.2472 & 0.4125 & 1.3298 & 1.2127 & 1.4470 \\
District of Columbia & 8 & 1.9101 & 0.4538 & 1.1819 & 0.8204 & 1.5435 \\
Florida         & 105 & 2.3152 & 1.4808 & 2.0194 & 1.7806 & 2.2583 \\
Georgia         & 33  & 2.3542 & 1.5691 & 2.1189 & 1.6624 & 2.5753 \\
Hawaii          & 5   & 2.4058 & 1.1220 & 1.7639 & 1.0852 & 2.4426 \\
Idaho           & 13  & 2.1801 & 1.1246 & 1.8311 & 1.0335 & 2.6286 \\
Illinois        & 67  & 2.2632 & 1.3351 & 1.8928 & 1.6040 & 2.1815 \\
Indiana         & 38  & 2.3518 & 1.2792 & 1.8571 & 1.5881 & 2.1261 \\
Iowa            & 24  & 2.1667 & 1.0788 & 1.7283 & 1.2274 & 2.2291 \\
Kansas          & 17  & 2.4039 & 1.4394 & 1.9824 & 1.5044 & 2.4603 \\
Kentucky        & 12  & 2.1597 & 1.1825 & 1.6711 & 1.1725 & 2.1696 \\
Louisiana       & 16  & 2.1480 & 1.2581 & 1.7031 & 1.2765 & 2.1297 \\
Maine           & 10  & 2.3044 & 1.3910 & 1.9607 & 1.2731 & 2.6483 \\
Maryland        & 21  & 2.2142 & 1.1205 & 1.7728 & 1.2382 & 2.3074 \\
Massachusetts   & 39  & 2.3466 & 0.9431 & 1.7233 & 1.4070 & 2.0396 \\
Michigan        & 51  & 2.1960 & 1.3165 & 1.8337 & 1.5217 & 2.1457 \\
Minnesota       & 25  & 2.4252 & 1.4620 & 2.1542 & 1.6321 & 2.6763 \\
Mississippi     & 5   & 2.3783 & 2.0000 & 2.2781 & 0.9182 & 3.6381 \\
Missouri        & 41  & 2.3160 & 1.4295 & 1.9827 & 1.6268 & 2.3387 \\
Montana         & 10  & 2.1225 & 1.4240 & 1.8202 & 1.0690 & 2.5715 \\
Nebraska        & 26  & 2.2278 & 1.0808 & 1.6977 & 1.3492 & 2.0463 \\
Nevada          & 11  & 2.3484 & 2.3745 & 2.4792 & 1.7692 & 3.1891 \\
New Hampshire   & 3   & 2.1882 & 1.4300 & 1.8091 & 1.6056 & 2.0126 \\
New Jersey      & 31  & 2.1925 & 1.4452 & 1.9661 & 1.5002 & 2.4320 \\
New Mexico      & 14  & 2.2762 & 1.5393 & 2.0750 & 1.3655 & 2.7845 \\
New York        & 71  & 2.1837 & 1.0558 & 1.6327 & 1.4244 & 1.8410 \\
North Carolina  & 55  & 2.2854 & 1.4275 & 1.9378 & 1.6379 & 2.2377 \\
North Dakota    & 3   & 2.4277 & 1.5400 & 1.9838 & 0.9936 & 2.9741 \\
Ohio            & 57  & 2.3874 & 1.2839 & 1.9111 & 1.5969 & 2.2253 \\
Oklahoma        & 658 & 2.3456 & 1.1328 & 1.8324 & 1.7468 & 1.9179 \\
Oregon          & 20  & 2.2713 & 0.8440 & 1.6130 & 1.2306 & 1.9954 \\
Pennsylvania    & 666 & 2.2713 & 1.0643 & 1.7417 & 1.6594 & 1.8240 \\
Rhode Island    & 6   & 2.3661 & 2.3817 & 2.7535 & 1.2373 & 4.2697 \\
South Carolina  & 24  & 2.2538 & 1.2692 & 1.8270 & 1.3790 & 2.2750 \\
South Dakota    & 5   & 2.3018 & 0.4620 & 1.3819 & 1.2523 & 1.5115 \\
Tennessee       & 29  & 2.4633 & 1.2490 & 2.0675 & 1.5640 & 2.5710 \\
Texas           & 104 & 2.2039 & 1.1030 & 1.7012 & 1.5040 & 1.8984 \\
Utah            & 12  & 2.5477 & 1.5483 & 2.1577 & 1.4140 & 2.9014 \\
Vermont         & 465 & 2.3695 & 1.1945 & 1.8394 & 1.7480 & 1.9308 \\
Virginia        & 36  & 2.2987 & 0.9575 & 1.7331 & 1.3814 & 2.0848 \\
Washington      & 671 & 2.3105 & 1.2937 & 1.8845 & 1.7982 & 1.9708 \\
West Virginia   & 10  & 2.1353 & 0.5610 & 1.3481 & 1.0176 & 1.6787 \\
Wisconsin       & 48  & 2.2806 & 0.6758 & 1.5271 & 1.2690 & 1.7851 \\
Wyoming         & 5   & 2.5281 & 1.0560 & 1.7921 & 1.0546 & 2.5296 \\ \bottomrule
\end{tabular}
\end{table*}